\documentclass[twocolumn]{aastex631}

\usepackage{mathtools}

\usepackage{color,soul}
\usepackage{subfigure}

\shorttitle{Radio emission}
\shortauthors{Tolman et al.}
\graphicspath{{./}{figures/}}

\begin{document}

\title{Electric field screening in pair discharges and generation of pulsar radio emission}

\author[0000-0002-2642-064X]{Elizabeth A. Tolman}
\affiliation{Institute for Advanced Study, 1 Einstein Drive, Princeton, NJ, 08540, USA}
\correspondingauthor{Elizabeth A. Tolman}
\email{tolman@ias.edu}

\author[0000-0001-7801-0362]{A.A. Philippov}
\affiliation{Center for Computational Astrophysics, Flatiron Institute, 162 Fifth Avenue, New York, NY, 10010, USA}
\affil{Department of Physics, University of Maryland, College Park, MD, 20742, USA}

\author[0000-0002-0067-1272]{A.N. Timokhin}
\affiliation{Janusz Gil Institute of Astronomy, University of Zielona Góra, ul. Szafrana 2, 65516, Zielona Góra, Poland}



\begin{abstract}
Pulsar radio emission may be generated in pair discharges which fill the pulsar magnetosphere with plasma as an accelerating electric field is screened by freshly created pairs. In this Letter we develop a simplified analytic theory for the screening of the electric field in these pair discharges and use it to estimate total radio luminosity and spectrum. The discharge has three stages.  First, the electric field is screened for the first time and starts to oscillate. Next, a nonlinear phase occurs. In this phase, the amplitude of the electric field experiences strong damping because the field  dramatically changes the momenta of newly created pairs. This strong damping ceases, and the system enters a final linear phase, when the electric field can no longer dramatically change pair momenta. Applied to pulsars, this theory may explain several aspects of radio emission, including the observed luminosity, $L_{\rm{rad}} \sim 10^{28} \, \rm{erg} \, \rm{s}^{-1}$, and the observed spectrum, $S_\omega \sim \omega^{-1.4 \pm 1.0} $.

\end{abstract}



\section{Introduction} \label{sec:intro}
Pulsars are rapidly rotating, highly magnetized neutron stars, most of which exhibit beamed radio emission. Several characteristics of this emission remain unexplained. Notably, the pulsar radio luminosity $L_{\rm{rad}}$ has rough magnitude $L_{\rm{rad}} \sim 10^{-7} - 10^{-2} \, L_{\rm{sd}}$, with $L_{\rm{sd}}$ the spindown luminosity, but has negligible dependence on $L_{\rm{sd}}$ \citep{lorimer2004handbook, szary2014radio}. Furthermore, the radio spectrum is $S_\omega \sim \omega^{-1.4 \pm 1.0} $, with $\omega$ the radiation frequency and $S_\omega$ the intensity emitted at that frequency \citep{bates2013pulsar}.

Compelling observational evidence suggests that in most pulsars radio emission is produced in the polar regions of the neutron star, where the star's rotation induces a strong electric field that extracts electrons from the neutron star surface and accelerates them to high energies \citep{sturrock1971model,ruderman1975theory}.  These electrons stream along curved magnetic field lines, radiating gamma rays which are absorbed by the pulsar magnetic field, producing electron-positron pairs. The resulting cascade, which includes additional contributions from inverse Compton scattering and synchrotron radiation, leads to the rapid production of a pair plasma in the polar cap region which screens out the initial inductive electric field~\citep{tsai1974photon,cheng1977pair,levinson2005large,timokhin2015polar,cruz2021kinetic}. 
Some process occurring in the pair plasma created in this discharge is likely to produce the pulsar's radio emission \citep{melrose2021pulsar}.

Recently, attention has turned towards the possibility that radio emission could be directly produced by the oscillating electric field that is created as the pair discharge screens the initial electric field \citep{beloborodov2008polar,timokhin2010time,timokhin2013current,melrose2020rotation,sasha}.  Specifically, direct numerical simulations \citep{timokhin2010time,timokhin2013current} of polar cap pair discharges have demonstrated that pair creation is a non-stationary, repeating process in which the accelerating electric field is screened by freshly created pair plasma, which then leaves the polar cap, causing the field to emerge again. \cite{sasha} showed that if pair creation is nonuniform across magnetic field lines, as it will be in any realistic pair cascade, a robust physical mechanism exists which generates coherent radio emission during the screening process. 
In this Letter, we study the highly relativistic, collisionless, and nonlinear plasma physics governing the evolution of this pair discharge's electric field amplitude and show that this plasma physics may explain the enigmatic pulsar radio luminosity, contribute to the spectrum, and give insight into other aspects of pulsar physics.  We verify our analytic models with direct kinetic simulations. 

In particular, during an initial nonlinear phase of the discharge, the electric field experiences strong damping which is roughly exponential. This strong damping ceases when the quiver of newly injected particles in the electric field is small.  Quantitative statement of the resulting condition on electric field amplitude yields an expression roughly consistent with the observed magnitude of $L_{\rm{rad}}$.  After this point, the discharge experiences a linear phase which only slightly decreases the emission amplitude but creates a relationship between emission amplitude and frequency which is consistent with the observed radio spectrum.

The Letter begins in section~\ref{sec:setup} by introducing the physical model and equations used to study the discharge plasma.  Then, section~\ref{sec:sim} describes the simulations of the discharge plasma which will be used to check the analytical work that forms the bulk of the paper. Section~\ref{sec:screen} describes the first screening stage of the discharge; section~\ref{sec:nonlinear} the following stage of nonlinear, strongly damped oscillations; section~\ref{sec:applin} the conditions that lead to the end of this stage's strong damping and set the overall emission amplitude; and section~\ref{sec:inter} the following linear stage which contributes to the pulsar spectrum. Finally, section~\ref{sec:conclusion} shows how our model explains observations of pulsar luminosity and may also give insight into the spectrum and other aspects of pulsar behavior.

\section{Description of model for discharge plasma physics}
\label{sec:background}
The discharge occurring in a pulsar polar cap has complex and uncertain spatial structure which inhibits analytic study. In this section, we describe the current understanding of the realistic polar discharge structure and relate it to the more simplified plasma physics problem we study in this Letter.  This will lead in section~\ref{sec:setup} to the statement of the equations studied in this paper.

The actual polar cap discharge proceeds through the cyclic production of clouds of pair plasma 
\citep{timokhin2010time,timokhin2013current,timokhin2015polar,sasha,cruz2021kinetic2}.  In this picture, an initially unshielded electric field in the pulsar's polar cap extracts electrons from the neutron star surface and accelerates them to significant Lorentz factors 
\begin{equation}
\label{eq:gamb}
\gamma_b \sim 10^7.
\end{equation}
Here, we have defined
\begin{equation}
    \gamma \equiv \frac{1}{\sqrt{1- \beta ^2}},
\end{equation}
with $\beta \equiv v/c$ ($v$ is the particle speed). These primary electrons radiate gamma rays which are eventually absorbed in the pulsar magnetic field, creating lower energy electron-positron pairs with an average Lorentz factor \citep{timokhin2015polar} 
\begin{equation}
\label{eq:gaml}
    \gamma_l \sim 10 - 10^3
\end{equation}
(here, the upper limit is more representative for pairs injected at earlier cascade stages, which are studied here)
and a thermal spread $T\sim m c^2$, with $m$ the electron mass \citep{arendt2002pair,timokhin2013current}. Though these pairs are much lower energy than the primary electrons, they move at essentially the same speed, such that the primary and secondary pairs form one cloud of plasma moving away from the star.  The secondary pairs interact strongly with the electric field, first shielding it, then setting up plasma waves caused by the overshoot of the shielding process. The primary population is barely affected by these plasma waves, such that it maintains a constant $\gamma_b \sim 10^7$ and its ability to create curvature radiation. Thus, there is a nearly constant source of new secondary pairs being added to the oscillations occurring in the pair plasma cloud, and the oscillations damp in response.  

Simulations of the two-dimensional polar cap region \citep{sasha} find that the created plasma waves are superluminal O-mode waves, which consist of an oscillating electric field with components along the background magnetic field and perpendicular to it, and an oscillating magnetic field perpendicular both to the background magnetic field and the oscillating electric field \citep{arons1986wave}.  In this Letter, we are concerned with understanding the physics governing the evolution of the amplitude of these waves.

The physical mechanism responsible for the generation of these waves works in at least two dimensions, requiring non-uniformity of pair creation across magnetic field lines. However, the electric field driving the waves is the oscillating parallel electric field, which interacts directly with the pair plasma through a plasma current.  This parallel electric field is determined by the dynamics of the discharge, i.e. by how the initial accelerating electric field is screened by injected pairs. Hence, wave evolution (at least in its initial stages) is determined by the evolution of the parallel electric field, which we can consider as a one dimensional problem. The waves observed in \cite{sasha} have small spatial frequencies $k$, so to allow analytic progress, we can also consider only temporal evolution of the electric field, i.e. we can ignore spatial variation along the magnetic field lines in both the initial electric field and in the particle injection rate.\footnote{We address some effects of a spatially non-uniform initial electric field in direct numerical simulations.}

Thus, in this Letter we model the discharge plasma physics with a one dimensional setup in which an initially strong electric field lies parallel to a magnetic field.  Low energy pairs with $\gamma_l$ are added to this system uniformly in space at a chosen rate, representing the pairs created in the actual plasma cloud (which, unlike our simplified one dimensional setup, has curved magnetic fields) by the higher energy pairs. In response, the electric field is shielded and begins to oscillate with spatial frequency $k=0$  (higher $k$ modes are not initially seeded because, unlike in \citet{sasha}, the setup does not have spatial structure). 

The physics governing the evolution of this oscillating electric field is expected to be roughly equivalent to the physics governing the low $k$ O-mode oscillations observed in \citet{sasha} and presumably also present in actual pulsar polar caps. (Indeed, as we discuss in section~\ref{sec:applin}, simulations of such an initially $k=0$ setup eventually fragment into higher-$k$ modes similar to those seen in two dimensional systems.) However, a $k=0$ setup is more tractable analytically and still gives significant insight into the physics of low-$k$ electromagnetic modes controlled by their parallel dynamics.

\section{Statement of studied equations and key quantities}
\label{sec:setup}
In this section, we present the specific equations that we use to understand polar cap physics through the picture introduced previously.

The pulsar polar cap region can be represented by a uniform, straight magnetic field $B_\star$ and a uniform parallel initial electric field $E_0$, corresponding to the magnetic field in the pulsar polar region and the unshielded electric field created there by the pulsar's rotation. These fields initially exist in a vacuum. For an aligned rotator, the electric field $E$ has initial strength  
\citep{deutsch1955electromagnetic,goldreich1969pulsar}
 
\begin{equation}
\label{eq:efield}
    E\left(t=0\right) \equiv  E_0 = -\frac{\Omega R_{\rm pc} }{c} B_\star,
\end{equation}
with $t$ time, $\Omega$ the pulsar angular velocity, $c$ the speed of light, and $R_{\rm pc}$ the polar cap radius, defined by 
\begin{equation}
\label{eq:rpc}
    R_{\rm pc} = R_\star \sqrt{ \frac{R_\star \Omega }{c}},
\end{equation}
where $R_\star$ is the neutron star radius, $R_\star \approx 1.2 \times 10^6 \, \rm{cm}$.

In an actual pulsar, the initial electric field $E_0$ extracts electrons from the neutron star into the polar cap vacuum and accelerates them to~\eqref{eq:gamb}, seeding a continual cascade of lower energy pairs of~\eqref{eq:gaml}. The high Lorentz factor $\gamma_b$ strongly suppresses the interaction of the high energy population with any plasma waves, so we neglect this population. Instead, we consider how~\eqref{eq:efield} evolves as lower energy pairs at $\gamma_l$ are added to the initial vacuum. The lower energy population will damp the electric field; the coupled evolution of the electric field and the lower energy pair population can be described by the relativistic Vlasov-Maxwell equations.

Specifically, the evolution of the electric field is governed by Amp\`ere's law, which for a curl-free magnetic field reads
\begin{equation}
\label{eq:ampere}
\partial_{t} E + 4 \pi {j}
 = 0,
\end{equation}
with $j$ the plasma current. The plasma consists of two species $s = +, -$: electrons (with charge $q_- = -e$), and positrons (with charge $q_+ = +e$). Both have mass $m$.   Furthermore, we define 
\begin{equation}
    u \equiv \gamma \beta
\end{equation}
as the particle momentum, equal to its component along the magnetic field.\footnote{In the strong pulsar magnetic field, any perpendicular momentum is rapidly radiated away, meaning that it is appropriate to consider only parallel momentum.} The electrons and positrons are each described by their corresponding distributions
\begin{equation}
    f_s(u,t),
\end{equation}
 so that the plasma current in~\eqref{eq:ampere} is given by
\begin{equation}
\label{eq:cur}
    j = c \sum_s q_s \int_{-\infty}^{\infty} \beta f_s\left(u,t\right)du,
\end{equation}
which couples Amp\`{e}re's law to the Vlasov equation.

The evolution of $f_s\left(u,t\right)$ is given by the Vlasov equation with a source term representing the injection of lower energy pairs:
\begin{equation}
\label{eq:vlasovtherm}
    \partial_{t} f_s + \frac{q_s E\left(t\right)}{mc} \partial_{u}{f}_s = \frac{\mathcal{S}  \left(u -u_l \right)}{2},
\end{equation}
Here, $u_l$ is the mean momentum of the newly injected lower-energy pairs
\begin{equation}
u_l \equiv \gamma_l \beta_l,
\end{equation}
with $\beta_l$ the value of $\beta$ corresponding to $\gamma_l$. The form of $\mathcal{S}$ characterizes the thermal spread of the newly injected pairs about their mean injection momentum; its integral gives the rate of increase of species density $n_s$ due to the addition of new pairs, i.e. $\partial_{t} n_s = \int_{-\infty}^{\infty} \left[\mathcal{S} \left( u -u_l \right)/2 \right] du$.  For our analytical arguments, we will take a simplified source model with all pairs injected at one velocity, i.e., 
\begin{equation}
\label{eq:vlasov}
    \partial_{t} f_s + \frac{q_s E\left(t\right)}{mc} \partial_{u} f_s = \frac{S\delta\left(u- u_l \right) }{2},
\end{equation}
with $S/2$ giving the rate per species of pair creation. (The effect of thermal spread is considered in appendix~\ref{sec:temp}.) For simplicity we assume this rate to be constant in time. 

Next, we apply a set of normalizations to Eqs.~\eqref{eq:ampere},~\eqref{eq:cur}, and~\eqref{eq:vlasov}. Let us define $t_0$ as the time it takes to change a pair's momentum by an amount $\Delta u=1$ in the initial electric field, i.e.
\begin{equation}
    t_0 \equiv \frac{mc}{\left|E_0\right|e},
\end{equation}
with $\left|E_0\right|$ the absolute value of the initial electric field.
Then, we normalize time to this value,
\begin{equation}
\label{eq:that}
    \hat{t} \equiv \frac{t}{t_0},
\end{equation}
 $f_s$ to the density injected in that time, 
\begin{equation}
    \hat{f}_s \equiv \frac{f_s}{St_0/2},
\end{equation}
and $E$ to its initial strength,
\begin{equation}
    \hat{E} \equiv \frac{E}{\left|E_0\right|}.
\end{equation}
(Later, when we wish to refer to the amplitude of an oscillating electric field, we will use $\overline{\hat{E}}$.)
Then,~\eqref{eq:vlasov} reduces to 
\begin{equation}
\label{eq:normv}
    \partial_{\hat{t}} \hat{f}_s + \frac{q_s}{e}\hat{E}\left(\hat{t}\right) \partial_u \hat{f}_s = \delta\left(u -u_l\right)
\end{equation}
and~\eqref{eq:ampere} to
\begin{equation}
\label{eq:normamp}
     \partial_{\hat{t}} \hat{E}\left( \hat{t} \right) + \hat{j} = 0,
\end{equation}
where~\eqref{eq:cur} becomes
\begin{equation}
\label{eq:j}
    \hat{j} \equiv \frac{1}{\xi} \sum_s \frac{q_s}{e} \int_{-\infty}^{\infty} \beta \hat{f}_s du. 
\end{equation}
Here, we have defined a parameter of key importance in determining the discharge dynamics: 
\begin{equation}
\label{eq:xidef}
\xi \equiv \frac{2 \left(E_0^2 /8\pi \right)}{m c^2 \left(S t_0 /2 \right)},
\end{equation}
which relates the energy in the initial pulsar electric field to the rest mass energy in the pairs injected during the time interval $t_0$. To find the appropriate value of this parameter to describe a pulsar, we introduce the Goldreich-Julian density \citep{goldreich1969pulsar}, $n_{{\rm GJ}} \equiv \left( \Omega B_\star \right) /\left( 2 \pi c e \right)$, and state that
\begin{equation}
    S\sim \frac{\lambda n_{{\rm GJ}}}{R_\star/c},
\end{equation}
with $\lambda$ the multiplicity. Then, the value of $\xi$ in a pulsar with magnetic field $B_\star$, multiplicity $\lambda$, and period $P$ is 
\begin{equation}
\label{eq:gammeq}
   \xi \sim 10^{12}  \left( \frac{B_\star}{10^{12} \, \textrm{G}}\right)^2  \left( \frac{10^5}{\lambda}\right)\left( \frac{1 \, \textrm{s}}{P}\right)^{7/2},
\end{equation}
a large value.

To analyze the nonlinear and linear stages of the electric field oscillations, it will be useful to combine~\eqref{eq:normv},~\eqref{eq:normamp}, and \eqref{eq:j} to obtain another relationship for the electric field. Noting that the current, represented by the second term in~\eqref{eq:normamp}, receives no contribution from particles newly injected at $u_l$, we can trivially use~\eqref{eq:normv}, without reference to the source term, to find the exact, nonlinear relationship
\begin{equation}
\label{eq:diffeq}
    \partial_{\hat{t}}^2 \hat{E} + \hat{\omega}^2\left(\hat{t}\right) \hat{E}\left(\hat{t} \right)  =0,
\end{equation}
where
\begin{equation}
\label{eq:freq}
   \hat{\omega}^2\left(\hat{t} \right) \equiv \frac{ \hat{n}_+}{\xi} \left< \frac{1}{\gamma^3}\right>_+ + \frac{ \hat{n}_-}{\xi} \left< \frac{1}{\gamma^3}\right>_-\,; 
\end{equation}
$\hat{\omega}$ is
the normalized relativistic plasma frequency characterizing $k=0$ oscillations  (which may be nonlinear or linear). \footnote{We note that for the case of a plasma where the electron and positron distributions are the same, this expression becomes, with normalization removed, $\omega^2 = \left( 8 \pi e^2 n_+ \left<1/\gamma^3 \right> \right)/m$, an expression helpful in obtaining~\eqref{eq:freqest}. }   
The normalized density $\hat{n}_s$ is defined by
\begin{equation}
\label{eq:dens}
   \int_{-\infty}^{\infty} du \hat{f}_s \equiv \hat{n}_s = \hat{t}   
\end{equation}
and the expected values $\left<1/\gamma^3\right>_+$, $\left<1/\gamma^3\right>_-$ by
\begin{equation}
    \int_{-\infty}^{\infty} \frac{du}{\gamma^3} \hat{f}_s \equiv \hat{n}_s \left< \frac{1}{\gamma^3} \right>_s. 
\end{equation}
In deriving~\eqref{eq:diffeq}, we have used the relation
\begin{equation}
\label{eq:relation}
    d \beta = \frac{du}{\gamma^3}
\end{equation}
and integrated by parts. The rest of the paper will consider the solutions to equations~\eqref{eq:normv},~\eqref{eq:normamp}, and~\eqref{eq:diffeq}, which describe the damping of the electric field in the polar cap by the newly injected pairs, and thus determine the radio emission characteristics. We emphasize that the relationships in this section are derived without linearization. They describe the oscillation of the plasma throughout all stages of the discharge, including the earliest stages, when the pair production rate is faster than the frequency $\hat{\omega}$.

\section{Direct numerical simulations}
\label{sec:sim}
The goal of the paper is the analytical study of the equations introduced previously, leading to relationships which can be used to characterize the radio emission produced in a pulsar polar cap discharge. To confirm these relationships, we compare them to kinetic particle-in-cell simulations with TRISTAN-MP v2 \citep{spitkovsky2005simulations,tristanv2}.

We initialize a one dimensional box with a uniform electric field and no plasma. At each subsequent timestep, pair plasma of momentum $u_l = 10$ [at the lower end of the range described by~\eqref{eq:gaml}] and small temperature $T = 0.1$ $m c^2$ 
(the effect of larger temperatures is considered in Appendix~\ref{sec:temp}) is injected into the box. The initial electric field amplitude and injection rate determine the value of $\xi$; as expected, the appropriately-normalized simulation results depend only on the value of $\xi$ and not on the electric field amplitude or source rate separately. (For convenience, after an initial test, only electric field strength was varied to vary $\xi$.) We run a set of simulations with values of $\xi$ ranging from $\xi = 2.8 \times 10^6$ to $\xi = 4.4 \times 10^7$.

After two time steps, there are ten skin depths in the simulation box, with each resolved by 700 cells. The simulation is continued, with density increasing and $\left<1/\gamma^3 \right>_{+,-}$ evolving, until the discharge is well past the transition to linearity; in all runs, the skin depth is resolved by at least 35 cells at the end of the simulation.

\section{Screening of initial electric field}
\label{sec:screen}
As pair creation begins in an unshielded electric field, the first event that occurs is the screening of the initial electric field, a process which has been considered previously in  \cite{levinson2005large} and \cite{cruz2021kinetic2}. In this section, we examine the physics of this stage, leading to an expression for screening time,~\eqref{eq:tscreennorm}, which agrees with previous results in the literature, and a new expression for the pair distribution function,~\eqref{eq:distscreen}.

The strong electric field initially present in the polar cap rapidly accelerates the first newly-injected lower energy pairs to large speeds, such that $\beta\rightarrow1$ and in~\eqref{eq:normamp} $\hat{j}$, defined in~\eqref{eq:j}, approaches 
\begin{equation}
\label{eq:approx}
   \hat{j} = \frac{2}{\xi} \int_{-\infty}^{\infty}\hat{f}_+ \left(u,\hat{t}\right) du = \frac{2 \hat{t}}{\xi},
\end{equation}
where in the second equality we have applied definition~\eqref{eq:dens}. Integration of~\eqref{eq:normamp} in time then gives that 
\begin{equation}
\label{eq:shieldE}
    \hat{E}\left(\hat{t} \right) = -1 + \frac{\hat{t}^2}{\xi}.
\end{equation}
The initial electric field is fully screened when $\hat{E} =0$, which occurs at time
\begin{equation}
\label{eq:tscreennorm}
    \hat{t}_{screen} = \xi^{1/2}.
\end{equation}

\begin{figure}
\centering
  \includegraphics[width=\columnwidth]{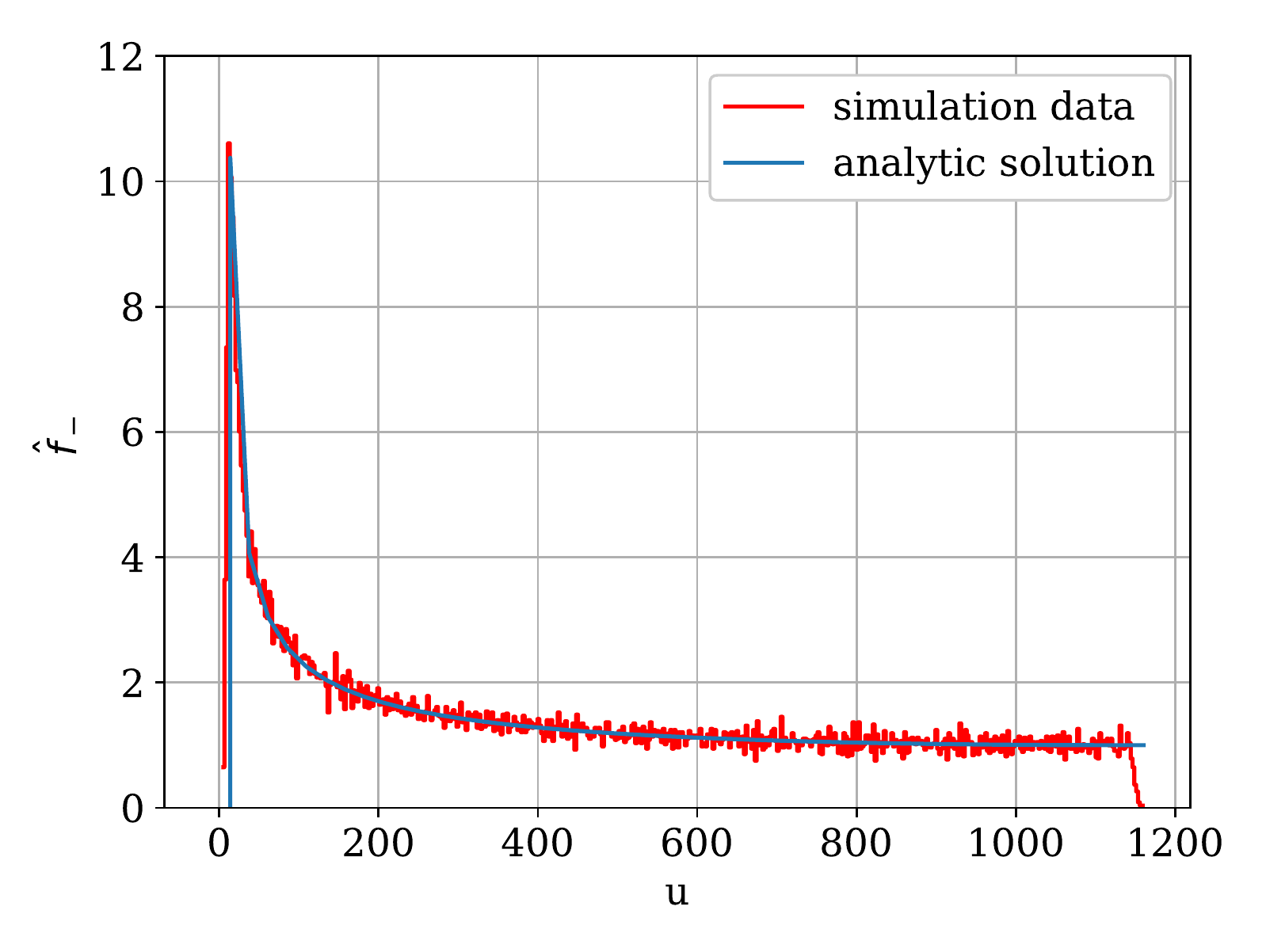}
  \caption{Electron distribution function at $\hat{t}_{screen}$ for $\xi = 2.8 \times 10^6$ from the analytic solution,~\eqref{eq:dist}, and from simulation.}
\label{fig:dist}
\end{figure}

As the electric field screens, particles are accelerated. Solving~\eqref{eq:normv} using~\eqref{eq:shieldE} shows the electron distribution function during the screening period is given by:
\begin{equation}
\label{eq:distscreen}
\hat{f}_-(u,\hat{t} ) = \begin{cases} 0, & u-u_l <0 \\
\frac{1}{1 - \hat{\tau}_{source}^2\left(u-u_l,\hat{t}\right)/\xi}, & 0 <u-u_l < \hat{t} - \frac{\hat{t}^3}{3 \xi}  \\
0, &  u-u_l >\hat{t} - \frac{\hat{t}^3}{3 \xi},
\end{cases}
\end{equation}
where the parameter $\hat{\tau}_{source}\left(u-u_l,t\right)$ is defined by the cubic equation
\begin{equation}
    u-u_l = \left(\hat{t}-\hat{\tau}_{source}\right) - \frac{1}{3\xi}\left(\hat{t}^3 -\hat{\tau}^3_{source} \right).
\end{equation}
The positron distribution function has the same form, but consists of negative values of $u-u_l$. We plot the analytic electron distribution function at the time of screening $\hat{t}_{screen}$ with the distribution function from a direct numerical simulation in Figure~\ref{fig:dist}, finding good agreement.
The largest disagreement with our analytic prediction occurs at low $u - u_l$, which corresponds to particles injected just before $\hat{t}_{screen}$, when the electric field is weak and~\eqref{eq:approx} begins to break down. The maximum $u$ obtained by the secondary pairs during the discharge is given by the maximum $u$ at which~\eqref{eq:dist} has support at $\hat{t}_{screen}$, i.e., $u-u_l = 2 \sqrt{\xi}/3$.  

At the time of screening, the electric field reaches zero, but the velocities obtained by the electrons and positrons continue to separate them, re\"establishing an electric field in the plasma and beginning the oscillations capable of creating radio emission, which we study next, beginning in the next section with study of the first, large oscillations.

\section{Nonlinear stage}
\label{sec:nonlinear}
The initial oscillations in the plasma following screening are, for large values of $\xi$, highly nonlinear and occur in a medium which changes quickly in comparison to the speed of the oscillation. As these oscillations proceed, they are damped by the continual creation of pairs, which are spun up into the wave oscillation, removing energy from the wave electric field. This section examines wave damping during this period.

The electric field amplitude in this stage evolves according to~\eqref{eq:diffeq}, which was derived without making any assumption of linearity. The behavior of the electric field is thus specified by the behavior of the frequency $\hat{\omega}^2$, defined in~\eqref{eq:freq}. Consideration of~\eqref{eq:freq} shows that $\hat{\omega}^2$ depends on density $\hat{n}_s$, which grows linearly as new pairs are added to the system, and on the values of $\left<1/\gamma^3\right>_s$, which evolve both due to the addition of new pairs at $u=u_l$ to the system and due to the large displacements in particle momentum that occur in response to a nonlinear wave. 
\begin{figure*}
\centering
\begin{subfigure}
\centering
\includegraphics[scale=0.5]{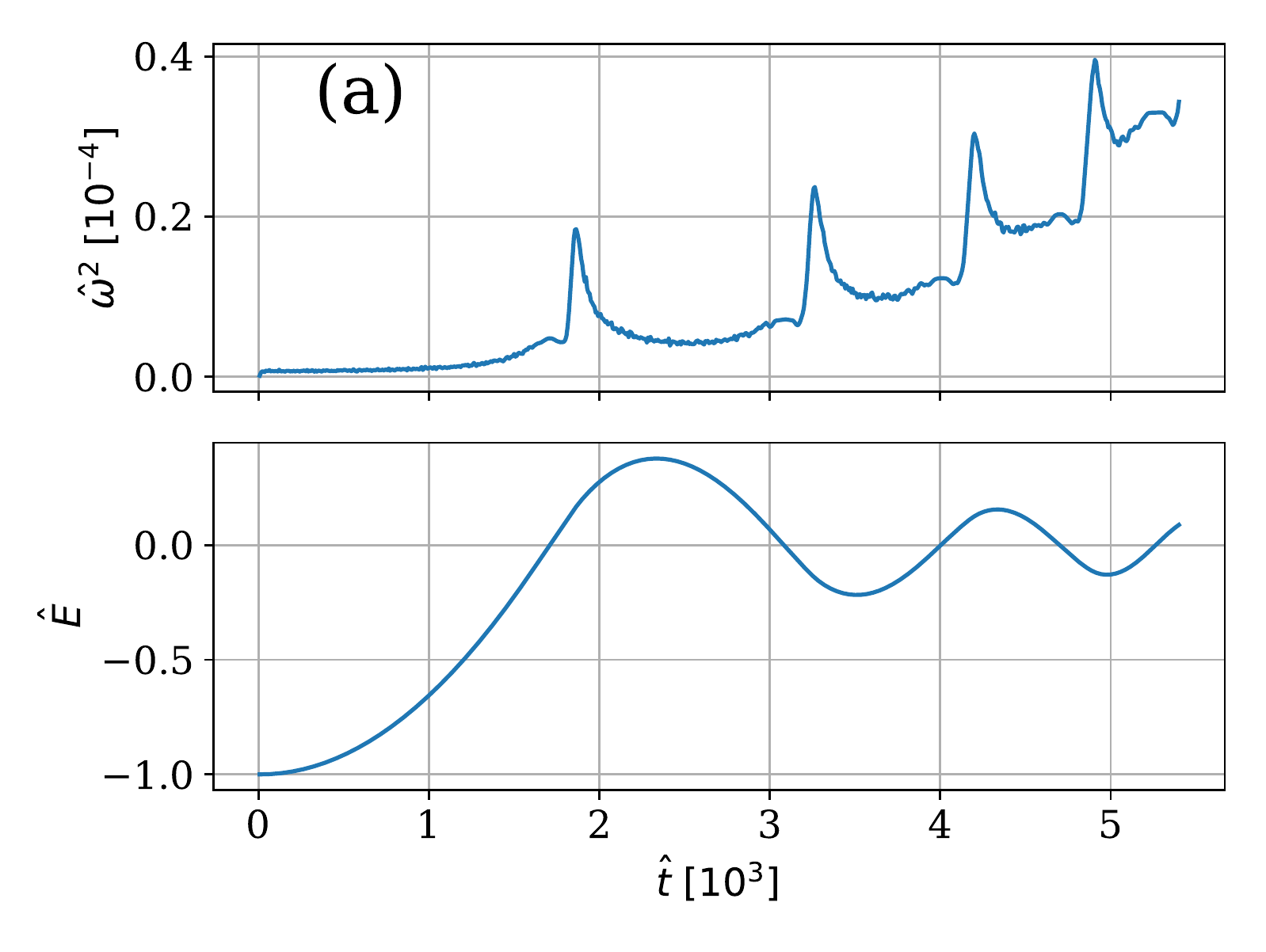}
\end{subfigure}%
\begin{subfigure}
\centering
\includegraphics[scale=0.5]{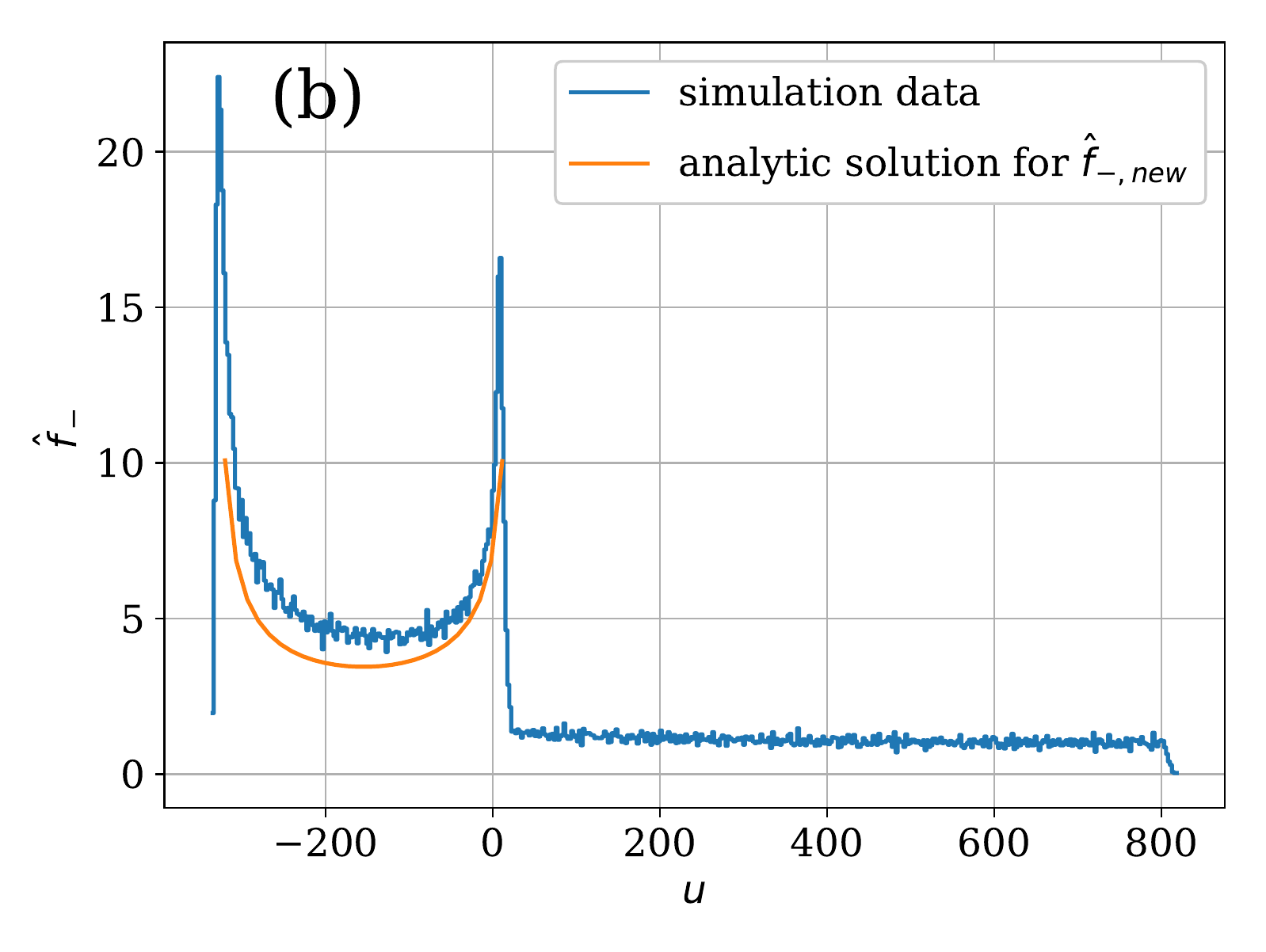}
\end{subfigure}
\begin{subfigure}
\centering
\includegraphics[scale=0.5]{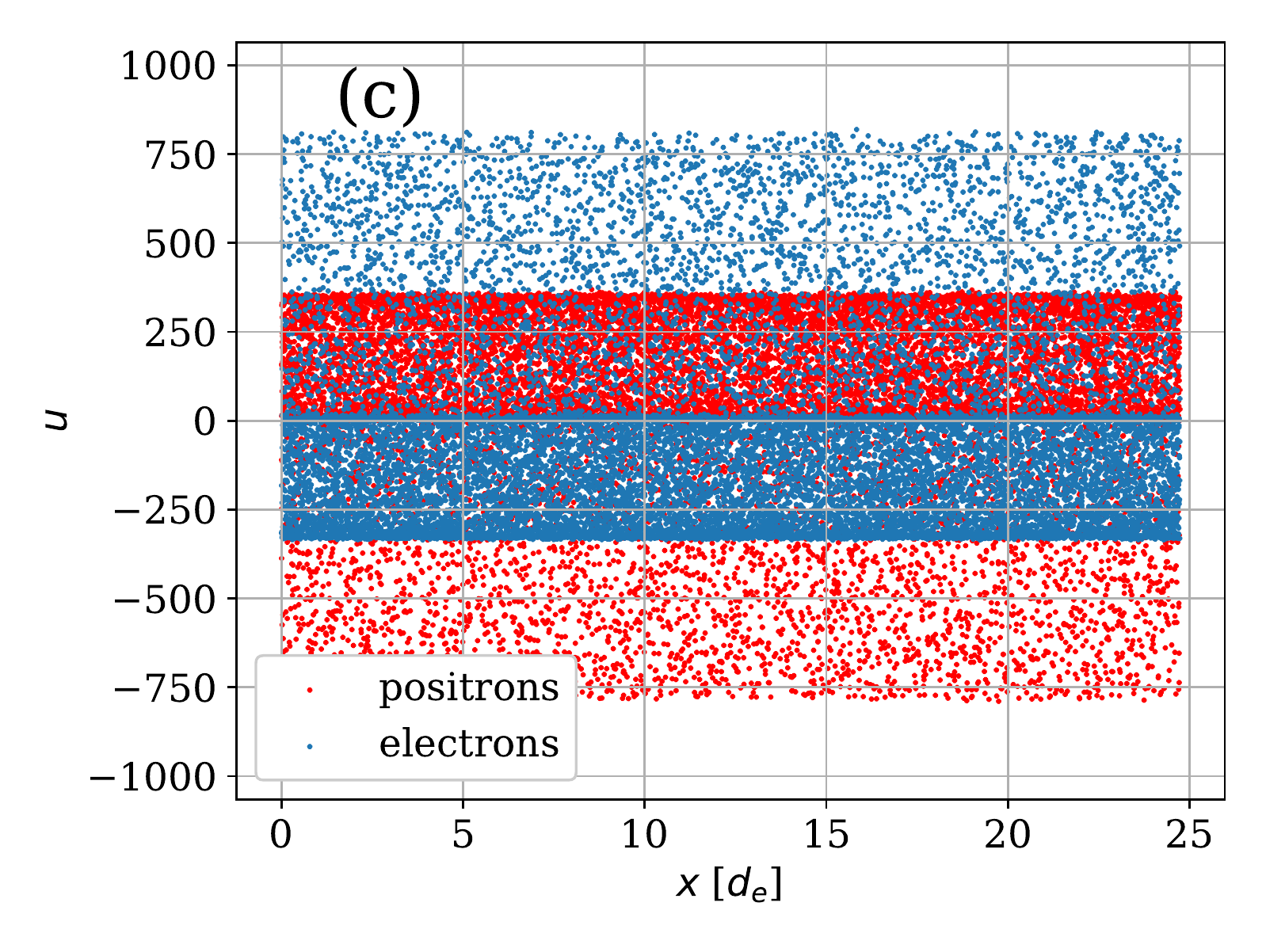}
\end{subfigure}
\begin{subfigure}
\centering
\includegraphics[scale=0.5]{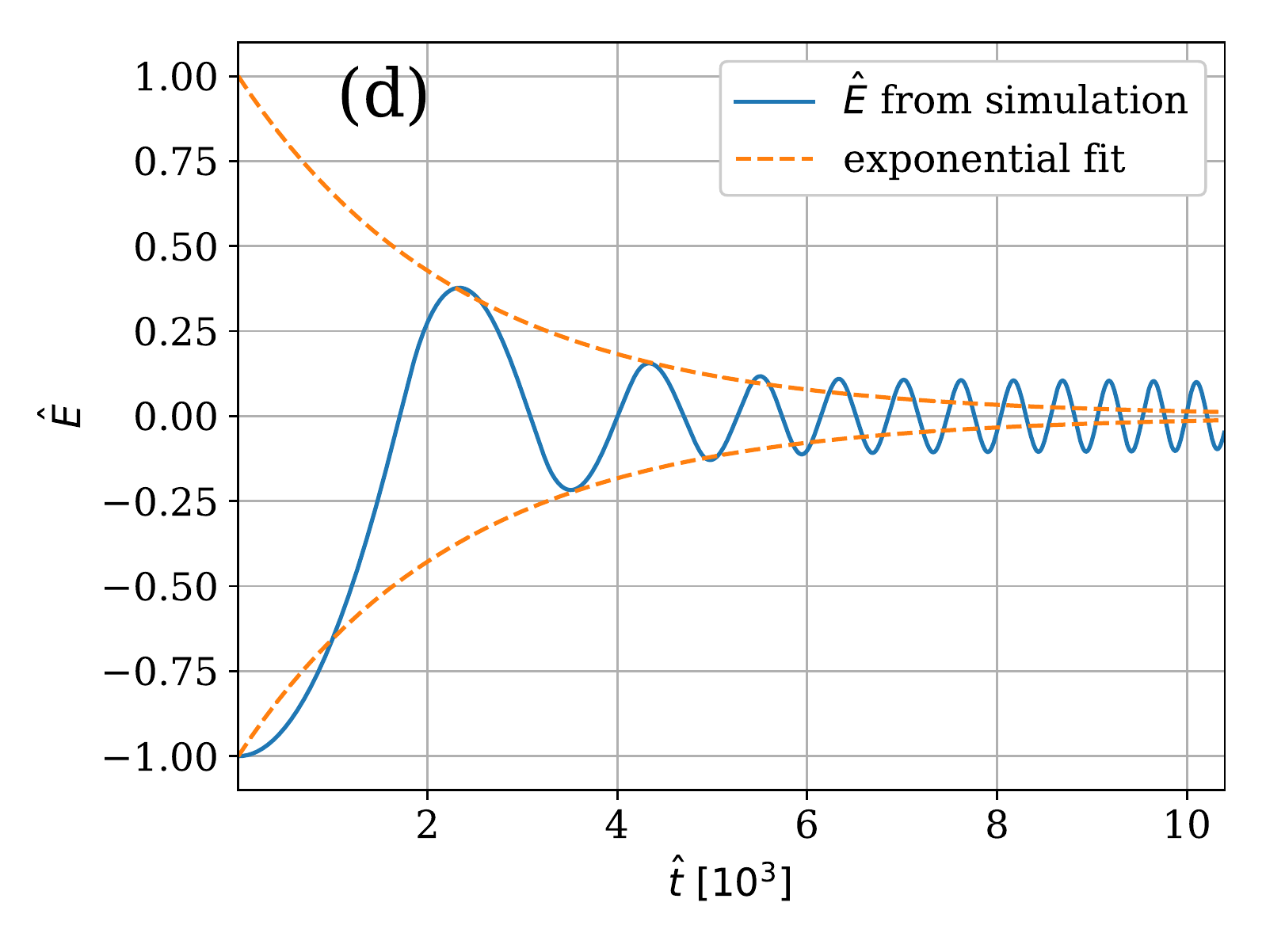}
\end{subfigure}
\begin{subfigure}
\centering
\includegraphics[scale=0.5]{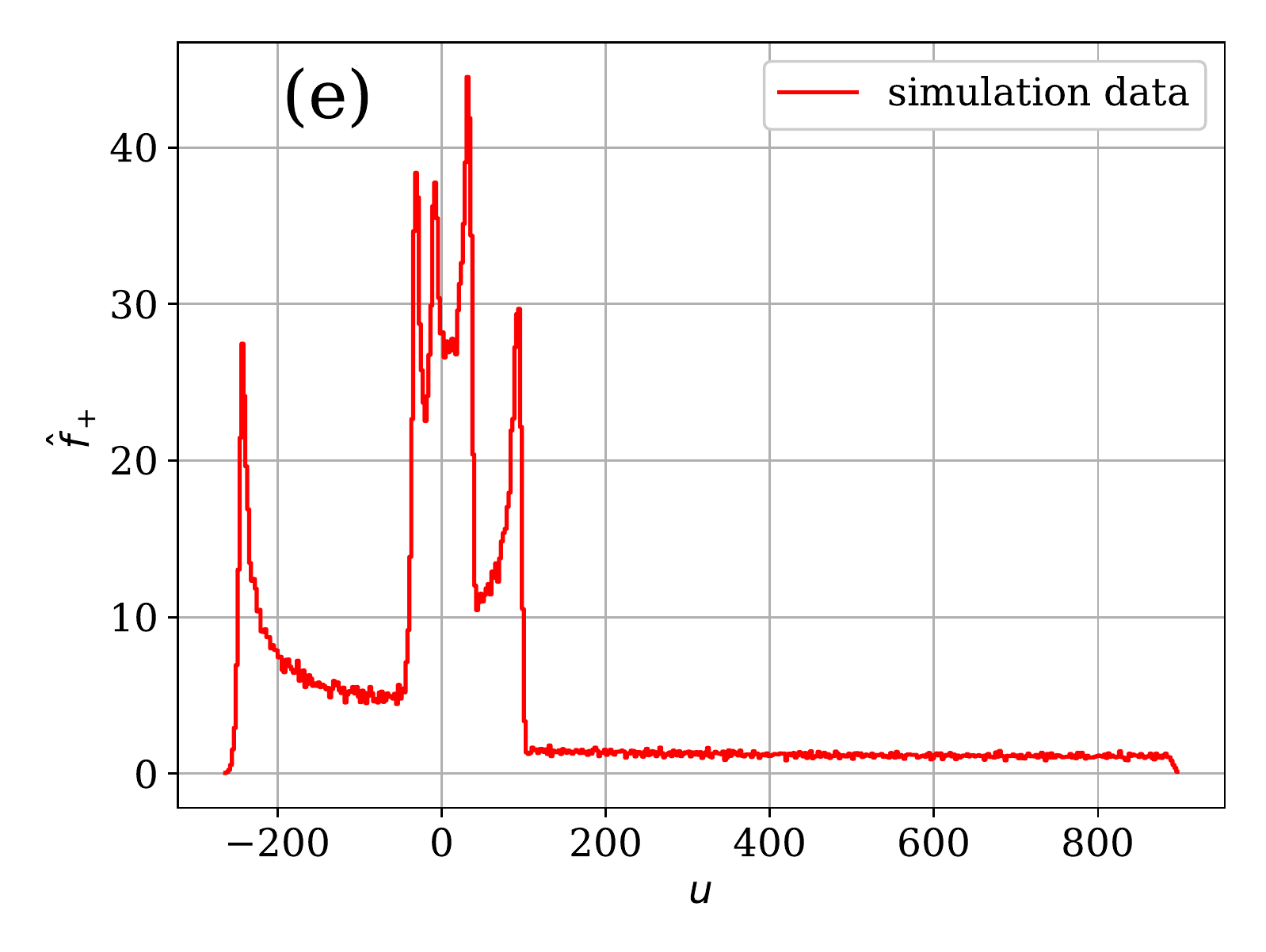}
\end{subfigure}
\begin{subfigure}
\centering
\includegraphics[scale=0.5]{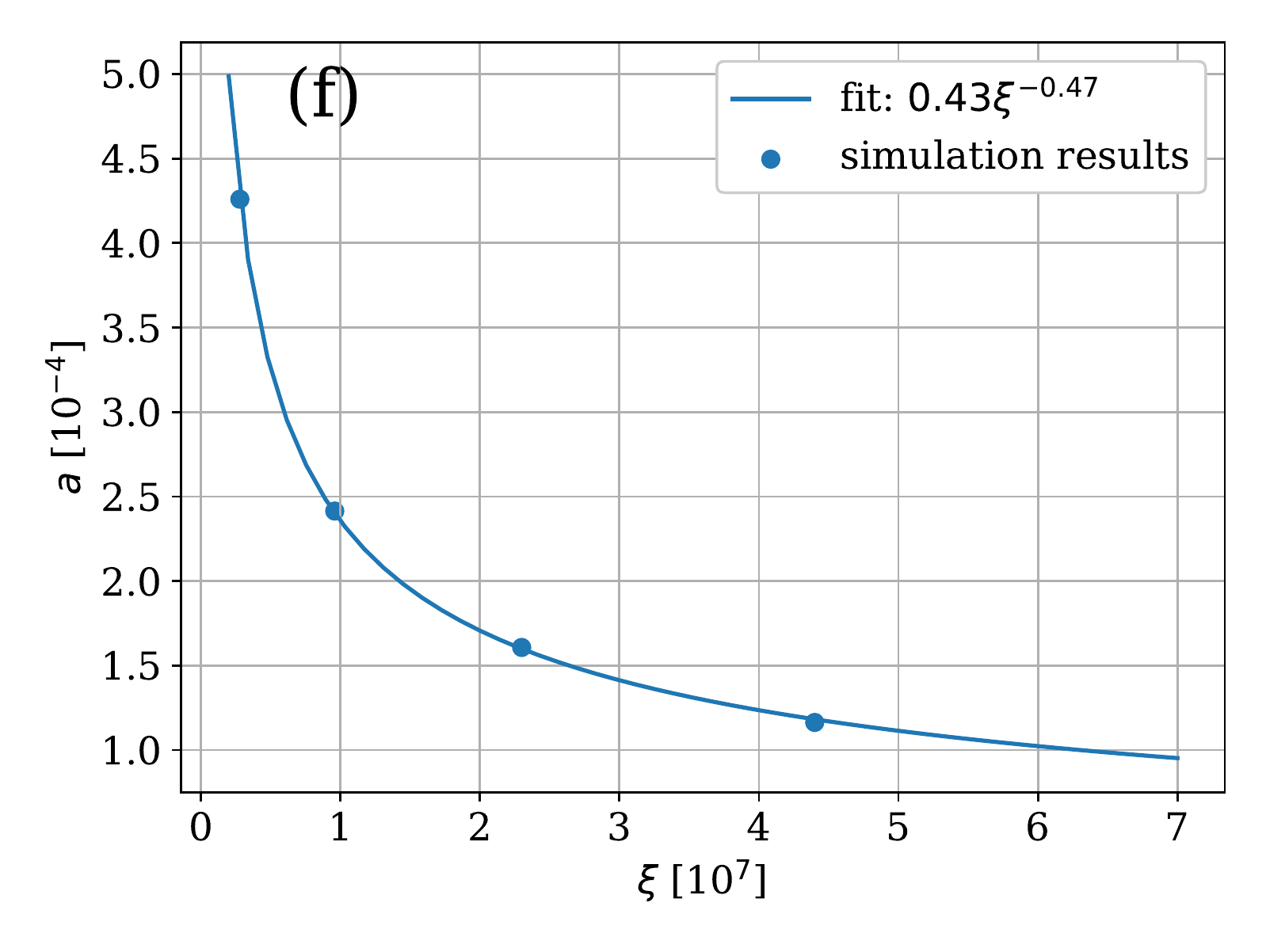}
\end{subfigure}
\caption{Quantities relevant to the nonlinear phase, all for a simulation with $\xi = 2.8 \times 10^6$: 2a) The early evolution of $\hat{\omega}^2$ and $\hat{E}$. The peaks in $\hat{\omega}^2$ occur later than the zeros of $\hat{E}$. 2b) Comparison of $\hat{f}_-$ at the second zero-crossing with the analytic model for $\hat{f}_{-,new}$ in~\eqref{eq:distnew}.  The value of $\hat{f}_-$ at negative $u$ is larger than that of $\hat{f}_{-,new}$ in part because $\hat{f}_-$ includes electrons already present at the time of screening, which are part of $\hat{f}_{-,old}$, that have backwashed to negative $u$. 2c) Phase space at the second zero-crossing, with every five hundredth particle marked. 2d) An exponential decay fit, $\overline{\hat{E}\left(\hat{t} \right)}=  e^{ -0.0016\hat{t}}$. The simulated $\hat{E}$ diverges from the fit near the transition to linearity, discussed in Section~\ref{sec:applin}. 2e) The value of $\hat{f}_+$ near the end of the nonlinear period, $\hat{t} = 5130$, showing several bounces stacked on top of each other. 2f) Fit of  the decay constant in $\overline{\hat{E}}\left(\hat{t} \right) =  e^{ -a\left(\hat{t}- \hat{t}_{screen} \right)}$ as found in the simulation and predicted by the fit presented in~\eqref{eq:actexp}. }
\label{fig:nonlinfig}
\end{figure*}

This evolution of $\hat{\omega}^2$ can be seen in the results of a simulation of the nonlinear stage, shown in the first panel of Figure~\ref{fig:nonlinfig}.  Here, $\hat{\omega}^2$ has large spikes, which represent times when the second derivative of the electric field increases and decreases dramatically in a short period of time. These spikes are caused by particles added at times when the electric field goes through zero.  These newly added pairs are not initially accelerated, so they contribute to a build up of a large number of particles near $u = u_l$.  Slightly later, one sign of pairs participating in this build up is dragged by the electric field through $u=0$, where the built up pairs contribute to a dramatic spike in $\left<1/\gamma^3 \right>$. Afterwards, as the particles are accelerated to high $u$ in the opposite direction, $\hat{\omega}^2$ decreases rapidly as pairs are accelerated to high $\gamma$.   Crucially, the peak of the spike occurs slightly after the zero of the electric field.

In~\eqref{eq:diffeq} a frequency which is described by this shifted spikes structure corresponds to efficient damping of the electric field.  Let us first consider an elementary demonstration of this effect; later, we will build a more involved model of the damping for the specific case of the polar discharge. 

Define time $\hat{t} =\hat{T}_h$ as a  point of zero-crossing for $\hat{E}$ that occurs at the end of half-period $h$ of oscillation. We endeavor to determine how a spiked frequency such as that observed in our simulation will change the derivative and amplitude of $\hat{E}$ between period $h = k$  and $h =k+1$.  Let us consider a simplified form of $\hat{\omega}^2$ containing only delta function spikes, offset by an amount $\Xi$ from $\hat{t} =\hat{T}_k$:
\begin{equation}
\label{eq:rough}
  \hat{\omega}^2 =  \delta \left(\hat{t} -\hat{T}_k - \Xi \right).
\end{equation}
Plugging this expression into~\eqref{eq:diffeq} and integrating the resulting equation over a small region $\hat{t}- \hat{T}_k \in \left[-\kappa, \kappa\right]$, where $\kappa > \Xi$, gives
\begin{equation}
\label{eq:damp}
    \partial_{\hat{t}}\hat{E} \left(\hat{T}_k +\kappa \right) -\partial_{\hat{t}}\hat{E} \left(\hat{T}_k-\kappa \right) = - \hat{E} \left(\hat{T}_k +\Xi\right).
\end{equation}
This equation represents a decrease in the magnitude of $\partial_{\hat{t}} \hat{E}$ between successive half-periods $h = k$ and $h = k+1$ of an electric field oscillation, causing the maximum amplitude of the electric field in the later half-period to be smaller, constituting damping of the electric field. The physical mechanism for this damping is the extraction of energy from the electric field to decelerate pairs in a way that causes a spike in the frequency.

This mechanism is at play in the nonlinear stage of the polar cap discharge.  To obtain more specific results, we now build a more involved model; this model includes study of the specific evolution of $\hat{f}_-$, yielding a more specific form for $\hat{\omega}^2$ than~\eqref{eq:rough} and, via~\eqref{eq:diffeq}, a functional form for electric field damping, which we will present in~\eqref{eq:exponential}. The treatment is imprecise due to the nonlinear physics involved.

 To develop a model for $\hat{f}_-$, let us call the amplitude of the electric field during half-period $h$ $\overline{\hat{E}}_h$; also, we consider the change in electric field amplitude and in $\hat{\omega}$ away from the zeros in $\hat{E}$ to be negligible. The electric field is approximately sinusoidal,\footnote{Note that sawtooth-shaped electric field oscillations, observed in \cite{levinson2005large}, do not occur in a kinetic treatment. We consider this point further in Appendix~\ref{sec:appsaw}.} so that in half-period $h$~\eqref{eq:normv} becomes
\begin{equation}
\label{eq:bouncev}
    \partial_{\hat{t}} \hat{f}_- + \left[\overline{\hat{E}}_h  \cos{\left(\hat{\omega} \hat{t}\right)} \right] \partial_u \hat{f}_- = \delta\left(u- u_l \right).
\end{equation}
Solving~\eqref{eq:bouncev} gives 
\begin{equation}
    \label{eq:dist}
    \hat{f}_- = \hat{f}_{-,new} \left(u-u_l, \hat{t} \right) + \hat{f}_{-,old} \left[u \hat{\omega} -  \overline{\hat{E}}_h \sin{ \left( \hat{\omega} \hat{t}\right)} \right].
\end{equation}
Here, $\hat{f}_{-,old}$, corresponding to the homogeneous solution of~\eqref{eq:bouncev}, represents the evolution of particles already present at the beginning of half-period $h$ of the oscillation. The term $\hat{f}_{-,new} \left(u, \hat{t} \right)$ gives the evolution of newly-added particles, and is given, in a half-period in which electrons are decelerated, by 
\begin{multline}
\label{eq:distnew}
\hat{f}_{-,new} (u-u_l,\hat{t} ) =
\\
\begin{cases} 0, & u-u_l > u_h \left(\hat{t} \right) \\
 \frac{1/ \overline{\hat{E}}_h }{\sqrt{g_+\left(u-u_l, \hat{t} \right)g_-\left(u-u_l, \hat{t} \right) }}, & u_h \left(\hat{t} \right)<u-u_l < 0 \\
0, &  u -u_l >0,
\end{cases}
\end{multline}
where
\begin{equation}
    g_+\left(u-u_l, \hat{t} \right) = 1 + \frac{\left(u-u_l\right)\hat{\omega}}{\overline{\hat{E}}_h}  - \sin{ \left( \hat{\omega} \hat{t} \right)},
\end{equation}
\begin{equation}
    g_- \left(u-u_l, \hat{t} \right) = 1- \frac{\left(u-u_l\right) \hat{\omega}}{\overline{\hat{E}}_h} + \sin{\left(\hat{\omega} \hat{t}\right)},
\end{equation}
and
\begin{equation}
\label{eq:umax}
    u_h \left(\hat{t}\right) \equiv \frac{\overline{\hat{E}}_h \sin{\left(\hat{\omega} \hat{t}\right)}- \overline{\hat{E}}_h}{\hat{\omega}} \leq 0.
\end{equation}
The second panel of Figure~\ref{fig:nonlinfig} shows a comparison between~\eqref{eq:distnew} and the distribution function at the second zero-crossing in the simulation illustrated in the first panel of Figure~\ref{fig:nonlinfig}, indicating rough agreement. For convenience, phase space in the simulation at this time is presented in the third panel of Figure~\ref{fig:nonlinfig}.

Let us consider how this distribution function changes the electric field between half-period $h=k$ and $h= k+1$; then, we will spread this change across a whole half period to obtain a continuous expression for the damping.

Consideration of~\eqref{eq:dist} reveals that the frequency~\eqref{eq:freq}, represented in our elementary demonstration by~\eqref{eq:rough}, is better represented near the transition between half-period $h=k$ and $h= k+1$ by
\begin{equation}
\label{eq:oscfreq}
    \hat{\omega}^2 \left( \hat{t} \right) = \frac{1}{\xi}\left[g\left(\hat{t}\right) + A \delta \left( \frac{\hat{t}-\hat{T}_k -\Xi}{\hat{t}_b} \right) \right].
\end{equation}
(Recall that $\hat{T}_k$ occurs at the end of half-period $k$.) Here, $g$ reflects the slow change in the frequency due to the evolving nature of $\hat{f}_{-,old}$ from one half-period to the next as the particles in each $\hat{f}_{-,new}$ enter the old particles in the next half-period.  This evolution includes a linear-in-time addition of density to the discharge and a slow increase in the overall value of $\left< 1 / \gamma^3 \right>$.  The second term models the spikes in $\left< 1 / \gamma^3 \right>$ caused by $\hat{f}_{-,new}$, which have characteristic height $A$ and width $\hat{t}_b$. In Appendix~\ref{sec:scaling}, we roughly estimate the scaling of these quantities with $\xi$:
\begin{equation}
    A\sim \xi^{0.2}
\end{equation}
and
\begin{equation}
    \hat{t}_b\sim \xi^{0.3}.
\end{equation}

Most damping is due to the delta function in~\eqref{eq:oscfreq} via the physics described in our elementary demonstration. As in~\eqref{eq:damp} we integrate over a small region $\hat{t}- \hat{T}_k \in \left[-\kappa, \kappa\right]$ to state (neglecting factors of order unity),
\begin{equation}
\label{eq:timedep}
    \partial_{\hat{t}} \hat{E}\left(\hat{T}_k+ \kappa\right) -\partial_{\hat{t}} \hat{E}\left(\hat{T}_k -\kappa\right) = - \xi^{-0.5} \hat{E}\left(\hat{T}_k +\Xi \right).
\end{equation}
The left side of the previous expression can be restated in terms of the frequency and the change in electric field amplitude:
\begin{equation}
     \partial_{\hat{t}} \hat{E}\left(\hat{T}_k+ \kappa\right) -\partial_{\hat{t}} \hat{E}\left(\hat{T}_k -\kappa\right) 
     =
    \hat{\omega} \overline{\hat{E}}_{k+1} - \hat{\omega} \overline{\hat{E}}_k.
\end{equation}
 (Recall that $\overline{\hat{E}}_{k+1}$ represents the wave amplitude in the next half-period of the oscillation.) Then, by representing the total change in amplitude as a rate multiplied by a time, we also state
\begin{equation}
\label{eq:intermediate}
     \hat{\omega} \overline{\hat{E}}_{k+1} - \hat{\omega} \overline{\hat{E}}_k 
     =
     \hat{\omega} \left [ \frac{\Delta \overline{\hat{E}}_h }{\Delta \hat{t}_h }\Delta \hat{t}_h   \right] ,
\end{equation}
with $\Delta \overline{\hat{E}}_h$ representing the change in electric field amplitude from one half-period to the next, and $\Delta \hat{t}_h$ the length of a half-period. In~\eqref{eq:intermediate}, we can make the replacement $\Delta \overline{\hat{E}}_h / \Delta \hat{t}_h \rightarrow d\overline{\hat{E}} / d\hat{t} $, with $ d\overline{\hat{E}} / d\hat{t} $ representing a continuous change in the amplitude of the electric field oscillations; furthermore, we note that $\Delta \hat{t}_h \sim 1/\hat{\omega}$ (again neglecting order unity factors). The continuous analog of~\eqref{eq:timedep}, that is, an expression which spreads the localized damping caused by a spiked frequency across time, is then:  
\begin{equation}
    \frac{d \overline{\hat{E}}}{d \hat{t}} = - \xi^{-0.5} \overline{\hat{E}}. 
\end{equation}
The evolution of the electric field amplitude during the nonlinear phase is then described by
\begin{equation}
\label{eq:exponential}
    \overline{\hat{E}}\left(\hat{t} \right) = e^{-\left(\xi^{-0.5}\right) \hat{t} },
\end{equation}
where the initial electric field amplitude is $\overline{\hat{E}} = 1$.  The initial period of electric field evolution during the polar cap discharge is thus marked by strong, exponential damping of the amplitude of the electric field oscillations. This is caused by particles added during zero-crossings of the electric field having their momenta reversed and accelerated to high values in the opposite direction, sapping energy from the electric field. This damping grows weaker with higher values of $\xi$, which correspond to weaker pair damping.

To verify this model, we use our set of simulations with several values of $\xi$ to fit an exponential decay $\overline{\hat{E}}\left(\hat{t} \right) =  e^{ -a\hat{t}}$ to the peaks of the first three peaks of electric field amplitude.  An example of this fit is shown in the fourth panel of Figure~\ref{fig:nonlinfig}. Then, we assemble the set of fit exponential decay constants $a$ and fit these constants to a power law function of $\xi$.  The resulting fit is 
\begin{equation}
\label{eq:actexp}
    \overline{\hat{E}}\left(\hat{t} \right) = e^{ -\left(0.4 \xi^{-0.5}\right) \hat{t}},
\end{equation}
in agreement with~\eqref{eq:exponential}. The fit for $a$ compared with the values of $a$ measured from simulation data can be seen in the final panel of Figure~\ref{fig:nonlinfig}.

\section{Transition to linearity}
\label{sec:applin}
As the discharge proceeds, the initial stage of slow, large-amplitude oscillations, which experience the exponential damping described by~\eqref{eq:exponential}, will cease, and the discharge will proceed to a phase of smaller oscillations at higher frequency. We call this later regime the ``linear" stage, which we analyze in Section~\ref{sec:inter}, revealing significantly weaker damping than in the nonlinear stage.\footnote{Oscillations with $k=0$, including those observed in our simulations before a late fragmentation into finite $k$ modes discussed in this section,  are exact nonlinear solutions to~\eqref{eq:diffeq} no matter their amplitude. For such a special wave, the techniques discussed in Section~\ref{sec:inter} require only that the timescale of background plasma variation be significantly longer than the period of oscillation. However, this condition is roughly equivalent to one requiring that the effect of the wave on particle motion be small, which is necessary for the linearization of the Vlasov equation.} In this section, we discuss the conditions that cause the transition between the nonlinear stage and the linear stage.

In the nonlinear stage, damping is caused by sharp spikes in $\hat{\omega}^2$.  These strong spikes will cease to exist when, in~\eqref{eq:diffeq},  
\begin{equation}
\label{eq:cond1prelim}
    \frac{\partial_{\hat{t}} \hat{\omega}^2} {\hat{\omega}^3} \sim 1,
\end{equation}
where $\hat{\omega}^2$, defined in~\eqref{eq:freq}, varies in time like~\eqref{eq:oscfreq}. The time variation in $\hat{\omega}$ comes from two sources: the change in the values of $\hat{n}_s$ and in the values of $\left< 1/\gamma^3 \right>_s$.  The component of $\partial_{\hat{t}} \hat{\omega}^2$ coming from the change in $\hat{n}_s$ is already small at the time of screening, and does not contribute to the spiking behavior.  Instead, the spiking results from large displacements in the particle momentum as the wave electric field grows from zero to its full amplitude over the course of a half-period of oscillation, changing the values of $\left< 1/\gamma^3 \right>_s$.

These displacements become negligible roughly at the point where the displacement in particle $u$ caused by the wave is insufficient to reverse the typical $u$ of the injected particles, $u_l$. The variation in a particle's momentum, $\delta u$, is found by considering the second term in~\eqref{eq:dist} or the maximum magnitude $u-u_l$ with support in~\eqref{eq:distnew},
\begin{equation}
    \delta u\sim \frac{\overline{\hat{E}}}{\hat{\omega}}.
\end{equation}
This quantity is a relativistic version of the quiver or jitter velocity familiar from classical plasma physics \citep{catto1989quiver}. That this quantity no longer be large enough to reverse $u_l$ requires
\begin{equation}
\label{eq:condprelim}
    \frac{\overline{\hat{E}}}{ \hat{\omega} u_l } \sim 1.
\end{equation}
After this point, spiking activity and the concomitant strong exponential damping considered in  Section~\ref{sec:nonlinear} cease.  We will show that the subsequent discharge evolution is characterized by much weaker damping, such that the final radio emission amplitude is roughly determined by the electric field amplitude at the time~\eqref{eq:cond1prelim} and~\eqref{eq:condprelim} are satisfied. 

Additional simplification is possible when stronger conditions than~\eqref{eq:cond1prelim} and~\eqref{eq:condprelim} hold, specifically,
\begin{equation}
\label{eq:cond1}
    \frac{\partial_{\hat{t}} \hat{\omega}^2} {\hat{\omega}^3} \ll 1,
\end{equation}
and
\begin{equation}
\label{eq:cond}
    \frac{\overline{\hat{E}}}{ \hat{\omega} u_l } << 1,
\end{equation}
which state, respectively, that the timescale of background plasma variation be significantly longer than the period of oscillation, and that the displacement in particle momentum caused by the wave be very small.

\begin{figure*}
\begin{subfigure}
\centering
\includegraphics[width=\columnwidth,trim= .5cm .5cm .5cm .5cm, clip=true]{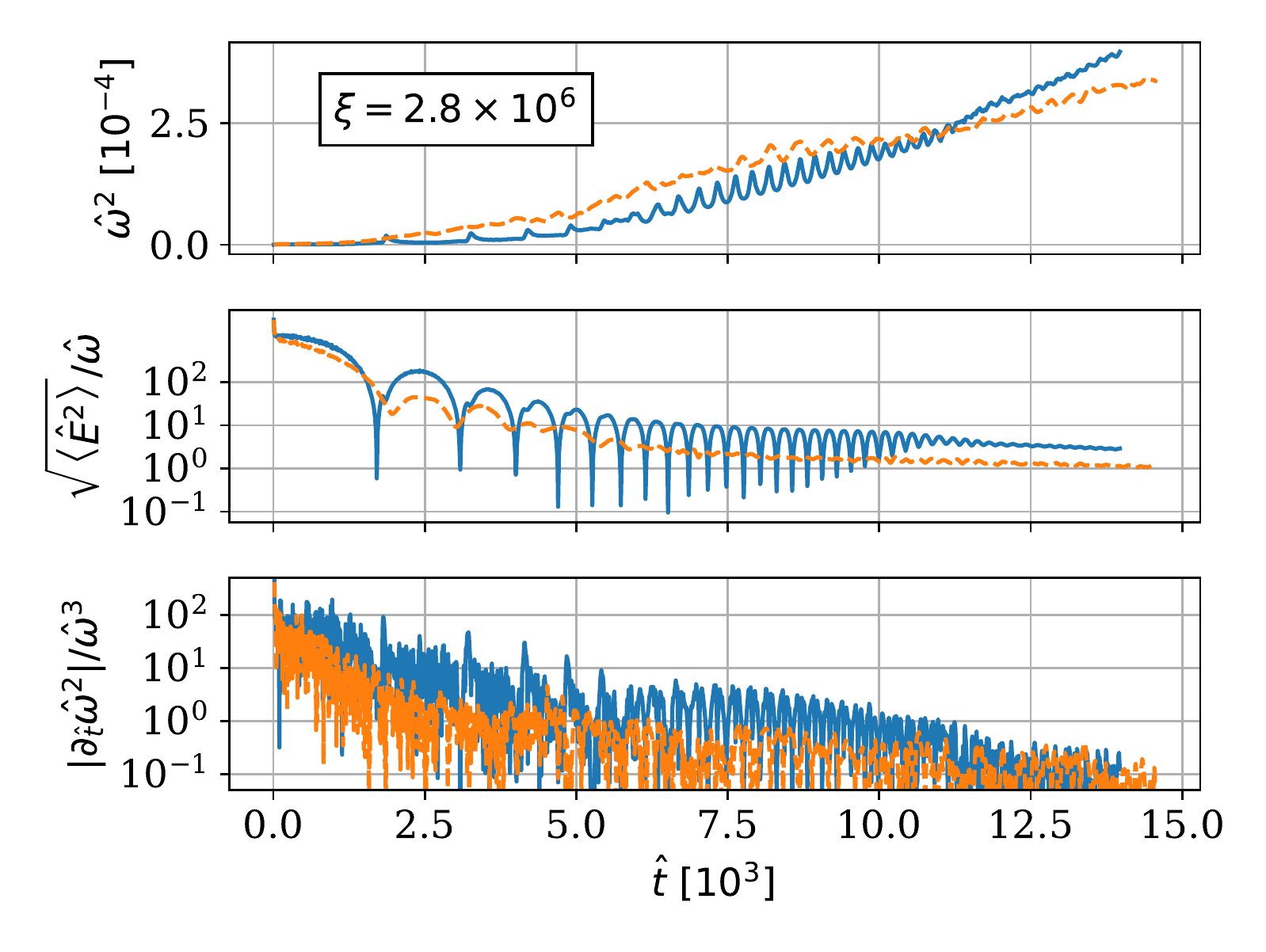}
\end{subfigure}%
\begin{subfigure}
\centering
\includegraphics[width=\columnwidth,trim= .5cm .5cm .5cm .5cm, clip=true]{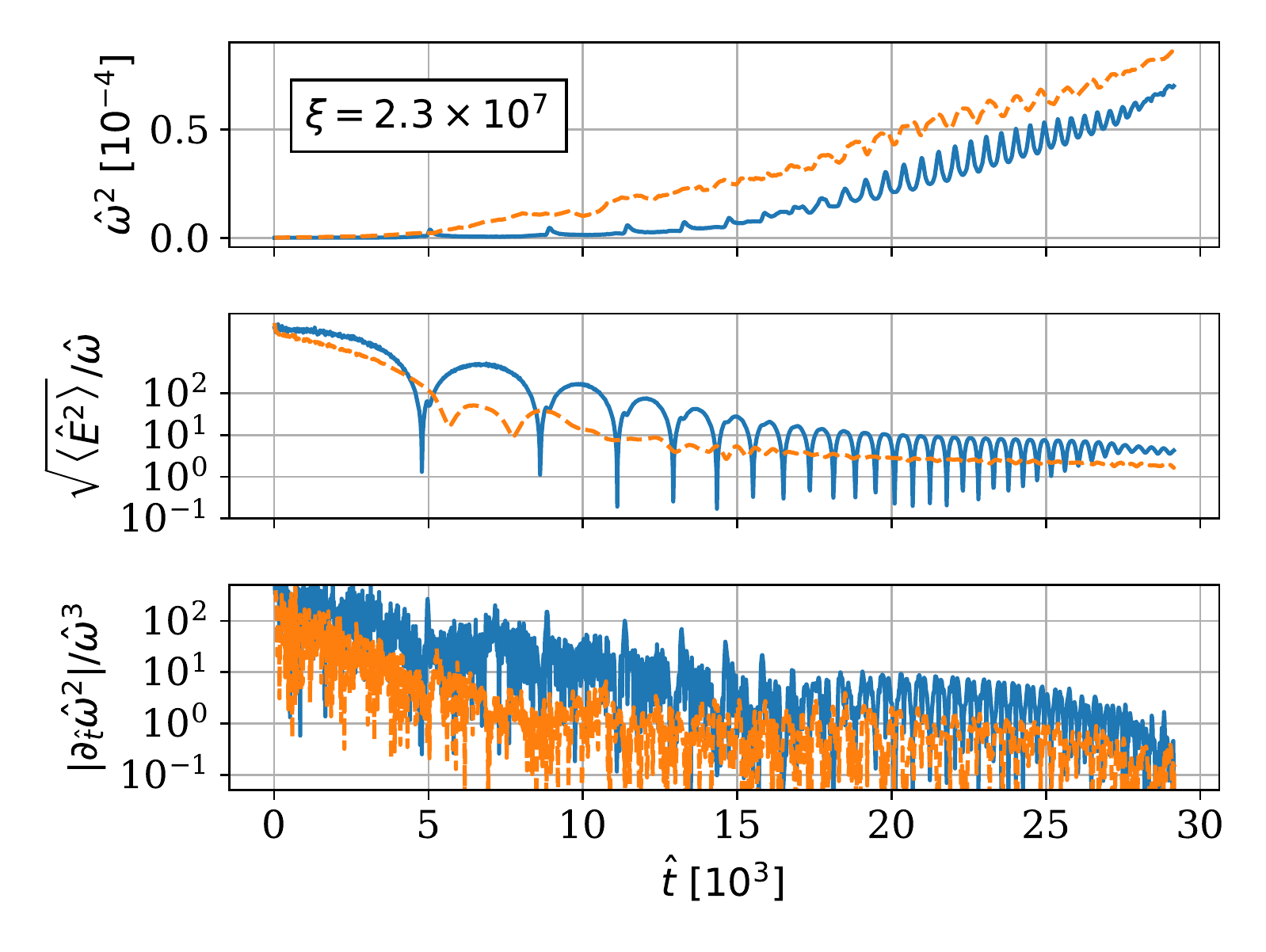}
\end{subfigure}
\caption{Evolution of parameters important to the transition to linearity in simulations with (left) $\xi = 2.8 \times 10^6$ and (right) $\xi = 2.3 \times 10^7$, shown in blue. Displayed are (from top to bottom) the squared frequency of the oscillation, the ratio of the root mean square amplitude of the electric field in the simulation box to the frequency (which characterizes the change in particle $u$ caused by the wave), and the normalized rate of change of the oscillation frequency. The dashed orange lines show the same quantities in a simulation run starting from an electric field with a finite $k$, as discussed at the end of Section~\ref{sec:applin}. The plotted frequency is the value of~\eqref{eq:freq} evaluated across the box, equivalent to the quantity plotted for the $k=0$ simulation, even though this simulation has spatial variation.}
\label{fig:approach}
\end{figure*}

We can observe the onset of~\eqref{eq:cond1prelim} and~\eqref{eq:condprelim}, and a later progression to~\eqref{eq:cond1} and~\eqref{eq:cond}, in our simulation. Figure~\ref{fig:approach} displays the values of $\hat{\omega}^2$, the root mean square value of the electric field in the simulation box divided by the frequency, $\sqrt{\left<\hat{E}^2\right>}/\hat{\omega}$, and $\partial_{\hat{t}} \hat{\omega}^2/ \hat{\omega}^3$ in two simulations.  
Considering the simulation with  $\xi = 2.8 \times 10^6$ (a higher value of $\xi$, which displays similar trends, is shown in the right panel of Figure~\ref{fig:approach}), we observe an initial stage where $\overline{\hat{E}}/\left(\hat{\omega} u_l \right)  >>1$, the plasma frequency changes quickly, and the electric field damps rapidly.  Once the simulation obeys~\eqref{eq:condprelim}, around $\hat{t} = 6000$, the strong damping ceases.  The frequency still has significant variation, resulting from coherent displacements in the plasma distribution function caused by an electric field that, though not able to reverse newly injected particles, is able to cause noticeable momentum changes. This means that although~\eqref{eq:cond1prelim} roughly holds,~\eqref{eq:cond1} is not satisfied. This situation is difficult to understand analytically, but corresponds in our simulation to negligible damping. 

\begin{figure}
\centering
\includegraphics[width=\columnwidth,trim= .5cm .5cm .5cm .5cm, clip=true]{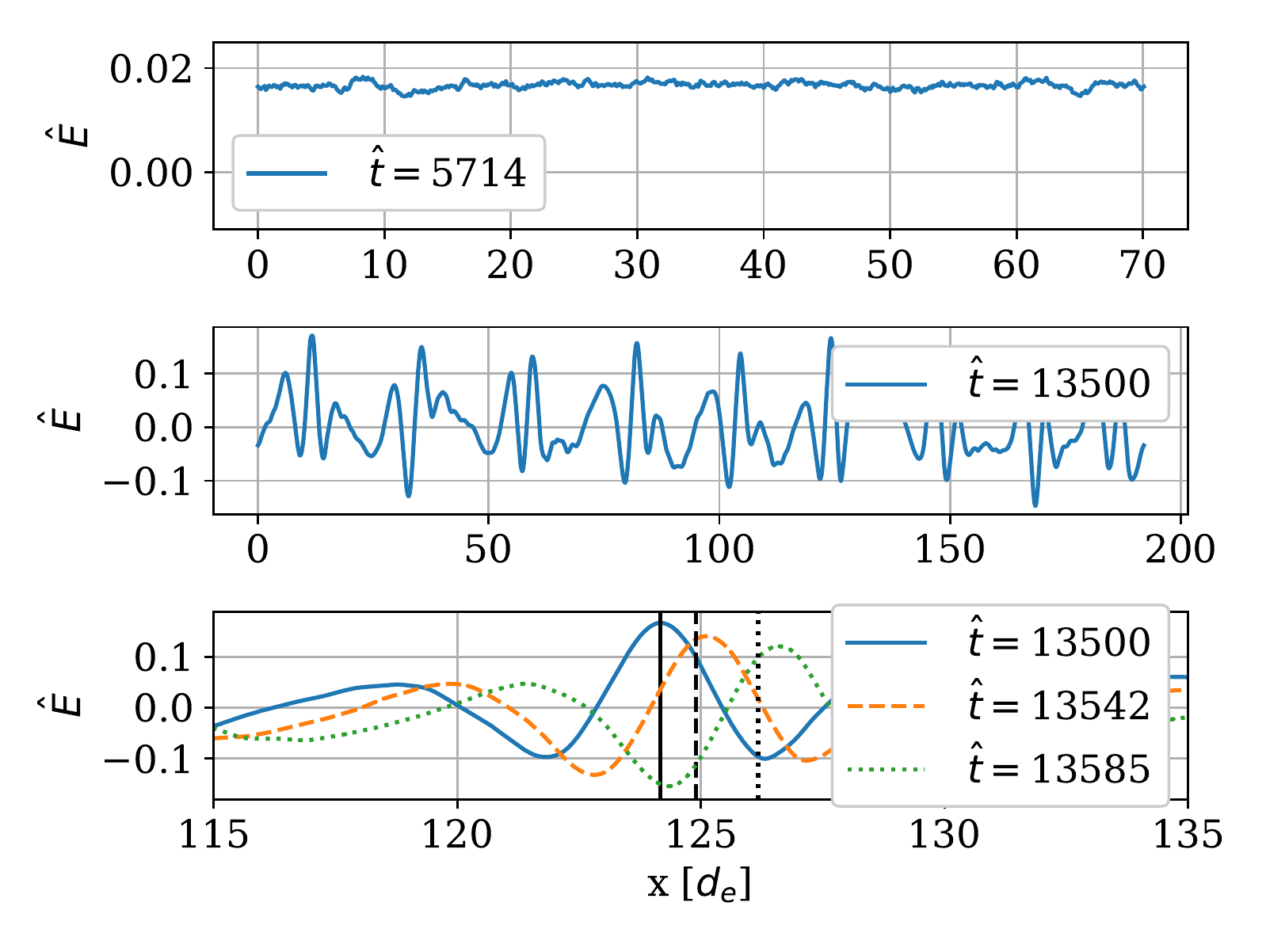}
\caption{Depiction of the fragmentation of the $\xi = 2.8 \times 10^6$ simulation into higher-$k$ modes. The first plot shows the electric field in the simulation box before the $k=0$ break up; the second shows the electric field after the fragmentation. Both of these plots show the full simulation domain, but the skin depth shrinks as time progresses, such that the axis labels of the two plots are different. The final plot zooms in on a small section of the box at three close times; the black lines move at the speed of light between the times and are overrun by the peak of one oscillation, indicating that the modes are superluminal. }
\label{fig:frag}
\end{figure}

Later, around $\hat{t}=11000$, our simulation of $k=0$ electric field oscillations fragments into superluminal finite-$k$ modes, which may be a result of a parametric-type instability like that considered in  \citet{cruz2021kinetic}. An illustration of this fragmentation is given in Figure~\ref{fig:frag}. After this point, the value of $\hat{\omega}^2$, where here $\hat{\omega}^2$ is given by~\eqref{eq:freq} evaluated across the simulation box, which is proportional to, but not the same as, the frequency of a given $k$ mode, does change slowly. This evolution occurs because the electric field is no longer able to cause coherent changes in the plasma distribution function. Conditions~\eqref{eq:cond1} and~\eqref{eq:cond} hold, and a new method of analysis, to be explained in Section~\ref{sec:inter}, becomes possible. 

We note here that we also ran simulations beginning with a finite $k$ mode. Specifically, we ran simulations starting from a sinusoidal electric field with one period in the simulation box (equivalent to a wave number $k = 0.63$ $d_e^{-1}$ with skin depth calculated from the plasma density two time steps into the simulation); the maximum initial electric field in these simulations is $\sqrt{2}$ larger than the uniform electric field in the normal simulations, such that the root mean square electric field amplitude is the same. Plasma is injected in the same way and at the same rate as in the $k=0$ simulations. The discharge in the nonzero-$k$ simulation quickly evolves into even higher-$k$ oscillations because the initial field is more quickly screened near the zeros of the electric field. There is no later fragmentation, and the oscillations remain superluminal throughout the discharge.

The resulting damping of these simulations is shown in Figure~\ref{fig:approach} by the orange dashed line. The nonzero-$k$ discharge exhibits a strong exponential damping phase, which ends when the root mean square amplitude of the electric field in the box approximately satisfies~\eqref{eq:condprelim}. This damping is followed by a weak damping phase, without the intermediate stage of minimal damping and $\overline{\hat{E}}/\left(u_l \hat{\omega}\right)\sim 1$ exhibited by the $k=0$ simulations. However, the value of~\eqref{eq:freq} evaluated across the simulation box in finite-$k$ simulations does not show the spiking behavior considered in Section~\ref{sec:nonlinear}, since the picture we advanced in that section will hold only locally.

\section{Linear stage}
\label{sec:inter}
In a plasma where~\eqref{eq:cond1} and~\eqref{eq:cond} are satisfied, the system enters a period in which the plasma wave is a small perturbation on the background and the frequency of the oscillation changes slowly, as quantified by~\eqref{eq:cond1}. In our simulations, this occurs after the fragmentation of the $k=0$ mode into finite $k$ oscillations; however, since the observed oscillations still have low values of $k$, the evolution of their amplitude can be roughly analyzed in the same way as a $k=0$ mode.

Specifically, the electric field differential equation,~\eqref{eq:diffeq}, can be analyzed directly using textbook WKB techniques~\citep{stix1992waves}, giving 
\begin{equation}
\label{eq:amplitude}
   \overline{\hat{E}}^2\sim \frac{1}{\hat{\omega}},
\end{equation}
with the electric field oscillation frequency determined by the current value of~\eqref{eq:freq}. In this stage, $\hat{\omega}^2$ changes due to increasing plasma density and evolving~$\left<1/\gamma^3 \right>_{+,-}$, which together yield an overall-increasing frequency, such that~\eqref{eq:amplitude} represents a slow damping of the electric field. [Neglecting the effect of changing $\left<1/\gamma^3\right>_{+,-}$,~\eqref{eq:amplitude} gives $\overline{\hat{E}}\sim 1/\hat{t}^{1/4}$, a dramatically weaker damping than~\eqref{eq:exponential}.]

Further physical insight into this damping can be obtained by arriving at~\eqref{eq:amplitude} via energy arguments. Using~\eqref{eq:normamp}, we can state 
\begin{equation}
\label{eq:1}
    \frac{1}{2} \partial_{\hat{t}} \hat{E}^2 = -\hat{j} \hat{E},
\end{equation}
which relates the change in the energy in the electric field to the power exerted on the current by the electric field. Via the same method used to obtain~\eqref{eq:diffeq}, we also have
\begin{equation}
\label{eq:2}
    \partial_{\hat{t}} \hat{j} = \hat{\omega}^2 \hat{E}
\end{equation}
with $\hat{\omega}$ defined in~\eqref{eq:freq}. Using~\eqref{eq:2} to substitute for $\hat{E}$ in~\eqref{eq:1}, we can arrive at
\begin{equation}
\label{eq:eq1}
    \partial_{\hat{t}} \left( \frac{\hat{E}^2}{2} + \frac{\hat{j}^2 }{2 \hat{\omega}^2 } \right) = -\frac{\hat{j}^2}{2 \hat{\omega}^2 } \left( \frac{1}{\hat{\omega}^2} \frac{\partial \hat{\omega}^2}{\partial \hat{t} }\right).
\end{equation}
The first term on the left side represents the wave energy in the electric field and the second the energy from the velocity of particles in the wave. When the wave frequency changes slowly, we can say these are roughly equal on average and replace~\eqref{eq:eq1} with an expression in terms of electric field amplitude:
\begin{equation}
    \partial_{\hat{t}} \overline{\hat{E}}^2 = -\frac{\overline{\hat{E}}^2}{2}\left( \frac{1}{\hat{\omega}^2} \frac{\partial \hat{\omega}^2}{\partial \hat{t} }\right).
\end{equation}
 From this equation, we can reproduce~\eqref{eq:amplitude}.
\begin{figure}
\centering
\includegraphics[width=\columnwidth]{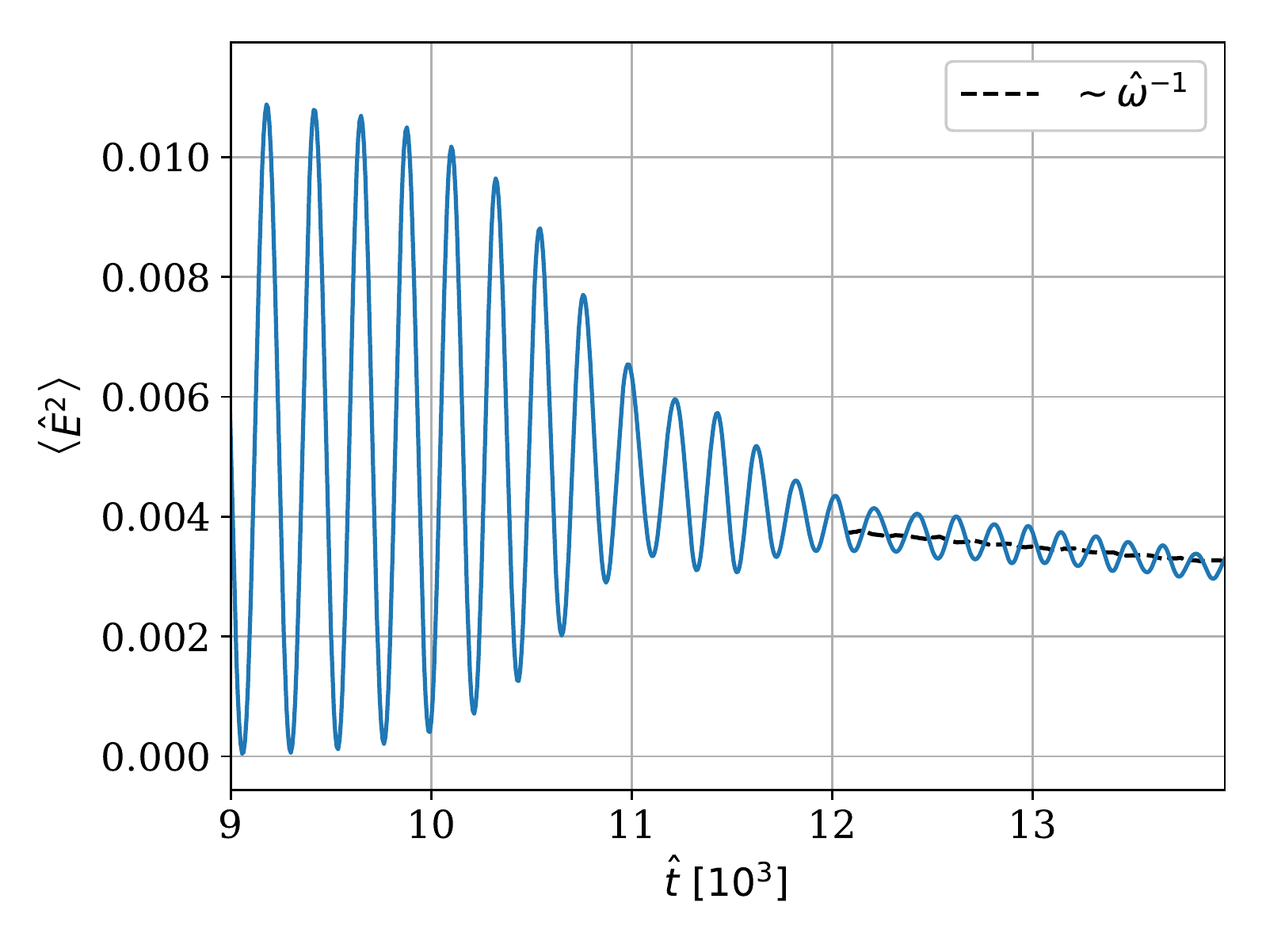}
\caption{The electric field in a simulation with $\xi = 2.8 \times 10^6$ compared to an  oscillation with amplitude~\eqref{eq:amplitude} and time-dependent frequency~\eqref{eq:freq}. The comparison begins after the simulation has fragmented into finite-$k$ modes, as discussed in Section~\ref{sec:applin}.}
\label{fig:linstage}
\end{figure}

The relationship~\eqref{eq:amplitude}, with a frequency calculated from the current value of~\eqref{eq:freq} across the whole simulation box, and the squared electric field strength given by its average value in the box, is compared to simulation in Figure~\ref{fig:linstage}, showing good agreement. 

In summary, the linear stage, which occurs late in the discharge, is marked by weak damping.  The nature of this damping is such that the relationship between electric field strength and frequency throughout the linear stage is given by~\eqref{eq:amplitude}.

\section{Conclusion with implications for pulsar radio emission}
\label{sec:conclusion}
We have developed a one-dimensional model for the evolution of wave amplitude during a pulsar polar cap discharge. In this model, the initial inductive electric field in the pulsar polar cap is screened by the creation of electron-positron pairs, setting up oscillations which continue to be damped by pair creation. This damping process has three phases: a screening phase, described in Section~\ref{sec:screen}, a nonlinear phase, described in Section~\ref{sec:nonlinear}, and a linear phase, described in Section~\ref{sec:inter}. The conditions for the transition between the nonlinear stage and the linear stage are considered in Section~\ref{sec:applin}.

In a realistic, two-dimensional pulsar polar cap, the discharge waves are initially electromagnetic \citep{sasha}; in addition, they will refract and may develop an even larger electromagnetic component as they also encounter cross-field gradients after their creation \citep{melrose2020rotation}. These electromagnetic waves eventually escape the plasma as Poynting flux.
Our one-dimensional model does not consider the magnetic component of the plasma oscillations, and cannot treat the escape process, which will be the focus of future work. However, the electromagnetic oscillations interact with the pair plasma through a parallel wave electric field and resulting plasma current, and this interaction, in which the physics of our model is  key, is responsible for the evolution of the amplitude of the radio emission. In this conclusion, we will thus use the physics understanding from our model to give new insight into multiple elements of pulsar observations. 

To start, we develop an expression for the frequency of the pulsar's emission. We note that in the linear phase of the discharge, the pair distribution function approaches a Maxwell-J\"uttner \citep{juttner1911maxwellsche} distribution drifting at $u_l$ (at the large Lorentz factors~\eqref{eq:gaml} characterizing the lower energy pairs, $\gamma_l \approx u_l$, and we will present results in terms of $u_l$).  For such a distribution, $\left<1/\gamma^3 \right>$ depends on $u_l$ and on the thermal spread of newly injected pairs, which is defined by the temperature, $T$. In particular, for pulsar conditions, where $ u_l $ $ \gg 1$, $T \gtrsim m c^2$ and $T/(mc^2) \ll$  $u_l$, the plasma frequency~\eqref{eq:freq} can be roughly represented by \citep{rafat2019wave} 
\begin{equation}
\label{eq:freqest}
    \omega \approx \sqrt{\left( \frac{8 \pi n_+ e^2}{m u_l^3}\right) \left( \frac{2 T}{mc^2}\right)}.
\end{equation}
The pulsar's radio emission is in the form of a superluminal O-mode, which has a frequency $\omega_O$ above the plasma frequency. Pair multiplicity in polar cap cascades of active pulsars relatively quickly reaches values of $\sim{}10^3$ and continues to grow until reaching the maximum multiplicity of $\sim\mbox{few}\times{}10^5$ \citep{timokhin2015polar,timokhin2019multiplicity}. Electromagnetic waves likely escape the plasma when it is characterized by a pair multiplicity between these values.%
\footnote{For a number of reasons, waves might escape the pair-forming blob, i.e. the region where the plasma density increases and the wave amplitude drops, before the pair multiplicity there saturates at $\kappa\sim10^5$.
The wave numbers $k$ of waves in our full discharge simulations \citep{sasha} are a non-negligible fraction of $\omega$/c, similar to the superluminal waves observed in the simulations which start with a non-zero $k$ discussed in this paper. Hence, the group velocity of oblique electromagnetic O-mode waves in hot plasma will be significant, and they should be able to outrun the region with increasing plasma density. 
The waves might also escape from the sides of the pair forming blob\textemdash the freshly created pair plasma is highly non-uniform, so the waves would refract towards regions of smaller density and might escape the blob (a scenario discussed in \citealt{melrose2021pulsar}). On the other hand, for some magnetic field lines, the pair-forming blob will be moving towards the NS  \citep[see][]{timokhin2013current} and the waves propagating towards the magnetosphere would not need to go through the region of increasing plasma density.}  
Evaluating~\eqref{eq:freqest} numerically for a pair multiplicity at the bottom of this range we get
\begin{multline}
\label{eq:obsfreq}
    \nu_O \geq 5 \, \rm MHz \left( \frac{n_+/n_{GJ}}{10^3}\right)^{1/2} \left( \frac{B_\star}{10^{12} \, \rm G}\right)^{1/2} 
    \\
    \times \left( \frac{P}{1 \, \rm s}\right)^{-1/2} \left( \frac{u_l}{10^3}\right)^{-3/2} \left( \frac{T}{mc^2} \right)^{1/2},
\end{multline}
with $\nu_O$ representing the frequency of O-mode waves.  Pulsar radio emission has been observed at frequencies as low as 15 MHz (see, for example, \citealt{pilia2016wide}), making~\eqref{eq:obsfreq} possibly consistent with observation. O-mode oscillations with higher wave numbers have higher frequencies; in addition, later in the discharge process, when the plasma multiplicity increases above $\sim 10^3$ and $u_l$ drops down to $\sim 10$, the frequency will increase. Both of these effects push~\eqref{eq:obsfreq} towards the radio range.

To continue, we present a set of arguments leading to an expression for pulsar radio luminosity. The luminosity depends on the amplitude of the emission when it leaves the pulsar.  As we described in Section~\ref{sec:nonlinear}, all strong damping of the plasma waves occurs in the nonlinear stage.  Considering~\eqref{eq:condprelim}, the end of the nonlinear phase occurs when
\begin{equation}
\label{eq:trans}
    \frac{e \overline{E}_{\star}}{ u_l m \omega_t c} \sim 1,
\end{equation}
where $\overline{E}_{\star}$ is the electrostatic component of the wave amplitude at the point of transition between the nonlinear and linear stages, $u_l \sim 10^3$ is the injection momentum of the lower energy pairs, 
and $\omega_t \sim  10^9 \, {\rm rad} \, {\rm s} ^{-1}$ defines the characteristic wave frequency at the end of the nonlinear phase.
Evaluation gives that $\overline{E}_{\star}=u_l m\omega_t c/e\sim 10^5 \,  \textrm{G}$. The time to reach the end of the nonlinear phase, $t_s$, can be found from~\eqref{eq:actexp}, $t_{s}\sim 2{\sqrt{\xi}} \ln(E_0/\overline{E}_{\star}) t_0$. Estimating $\ln(E_0/\overline{E}_{\star})\lesssim 10$ and using~\eqref{eq:xidef}, one obtains 
\begin{equation}
\label{eq:tau_screen}
t_{s} \lesssim 0.5 \left( \frac{P}{1 \, \rm s}\right)^{1/4} \left(\frac{R_{\rm{pc}}}{c}\right).
\end{equation}
This time is shorter than the light-crossing time across the polar cap, $R_{\rm{pc}}/c$, so the nonlinear phase of the discharge completes in all active pulsars.

As described in Section~\ref{sec:inter}, the damping of the electric field is weak during the linear stage, so the rough overall radio luminosity is likely to be set at the end of the nonlinear phase. (The end of the nonlinear phase also corresponds to the lowest frequencies of emission, because $u_l$ will decrease and multiplicity will increase in the linear phase. The pulsar spectrum declines with frequency, such that luminosity is dominantly determined by the low-frequency contributions occuring at the end of the nonlinear stage.) Emission across the pulsar polar cap gives a radio luminosity of $L_{{\rm rad}} \sim c \left(\overline{E}_{\star}\sin\alpha)^2 (\eta \pi R_{\rm pc}^2\right)/(4\pi)$, where $\alpha$ is the typical angle between the direction of wave propagation and the background magnetic field (see \citealt{sasha}) and $\eta$ characterizes the fraction of pair producing field lines in the polar cap. Using \eqref{eq:trans}, one obtains
\begin{multline}
\label{eq:lrad}
L_{{\rm rad}} \sim 10^{28}\textrm{erg}\, \textrm{s}^{-1}
\left(\frac{\omega_t}{10^9 \, \textrm{rad} \, \textrm{s}^{-1} }\right)^{2} \left(\frac{P}{1 \textrm{s} }\right)^{-1}
\\
\times
\left(\frac{u_l}{10^3}\right)^2 \eta \sin^2\alpha.
\end{multline}
This luminosity is towards the lower end of the range of observed radio luminosity \citep{szary2014radio,wu2020luminosity}; higher values of luminosity may be explained by the escape of some radiation before the end of the nonlinear stage. The prediction~\eqref{eq:lrad} can also be expressed as a fraction of the pulsar's spindown luminosity, $L_{{\rm sd}} = c E_0^2 \pi R_{\rm pc}^2/(4\pi)$ \citep{spit06}:
\begin{multline}
    \frac{L_{{\rm rad}}}{L_{{\rm sd}}} \sim \frac{\overline{E}_{\star}^2}{E_0^2}(\eta \sin^2\alpha) \sim 2\times 10^{-4}\left(\frac{P}{1 \, \textrm{s} }\right)^{3} 
    \times
    \\
    \left(\frac{B_\star}{10^{12} \, \textrm{G} }\right)^{-2} \left( \frac{\omega_t}{10^{9} \, \textrm{rad} \, \textrm{s}^{-1}} \right)^2 \left( \frac{u_l}{10^3}\right)^2(\eta \sin^2\alpha).
\end{multline}

Observations show practically no dependence of radio luminosity on pulsar parameters \citep{szary2014radio,wu2020luminosity}; however,~\eqref{eq:lrad} shows a non-negligible dependence on $B_*$ and $\Omega$. We should note that the uncertainties in our simplified estimate for $\omega_t$ and the numerical factors $\eta$ and $\alpha$ may substantially reduce this dependence. For example, in young energetic pulsars with large $\Omega$ (and $B_*$) the pair creation front is expected to be less inclined with respect to the background magnetic field, which reduces $\alpha$ and hence the dependence of radio luminosity on these parameters. We will refine these dependencies in our future multi-dimensional studies.

Next, although the overall radio luminosity is set at the transition between the nonlinear and linear phases, some modification in amplitude does occur in the linear phase. We expect that radiation will escape the plasma throughout the linear stage (and possibly during the nonlinear stage as well) in a continuous process that occurs concurrently with the generation of the plasma waves. The relationship between amplitude and frequency in the linear stage, given by~\eqref{eq:amplitude}, would thus be visible to an observer if, across the polar cap, the wave amplitude became unaffected by the plasma at a variety of different points in the linear damping process, when the value of the plasma frequency is different. This relationship roughly agrees with the observed spectrum of emission, $S_\omega \sim \omega^{-1.4 \pm 1.0} $, with  $S_\omega$ the intensity emitted at that frequency \citep{bates2013pulsar}. This suggests that the pulsar radio spectrum may result in part from the linear damping physics of the pair discharge.

Finally, the distribution of particles right after the end of the initial screening,~\eqref{eq:distscreen}, can be used as a model for the spectrum of particles flowing back from the discharge to the pulsar surface, allowing greater understanding of the X-ray hotspots observed by NICER \citep{gendreau2016neutron,salmi2020magnetospheric}.

\begin{acknowledgments}
E.T. is indebted to a helpful comment on Math StackExchange from user ``DinosaurEgg." This research made use of the ``Tristan-MP v2" particle-in-cell code. We thank Peter Catto, Lev Arzamasskiy, Carolyn Raithel, and Fabio Cruz for helpful conversations. Hayk Hakobyan provided expert advice on code usage. E.T. was supported by the W.M. Keck Foundation Fund at the Institute for Advanced Study. A.P. was supported by NSF grant no. PHY-2010145. A.T. was supported by the grant 2019/35/B/ST9/03013 of the Polish
National Science Centre.
\end{acknowledgments}
\appendix
\section{Discussion of sawtooth oscillations}
\label{sec:appsaw}
Works that study pulsar polar cap discharges using a fluid treatment observe sawtooth-shaped oscillations, where electric field evolution is linear except at turning points \citep{levinson2005large}.  There have been tentative suggestions in the literature that such sawteeth may be a generic feature of relativistic pair discharges, extending also to kinetic treatments \citep{cruz2021kinetic}.  In fact,  sawteeth do not occur in a kinetic treatment.

To understand why, let us first consider the origin of sawteeth in a fluid treatment. 
We can define a Lorentz factor associated with the electron and positron fluids as, with $s = +,-$ again indicating species,
\begin{equation}
    \gamma_s \equiv \frac{1}{\sqrt{1-\frac{v_s^2}{c^2}}},
\end{equation}
with $v_s$ the velocity of species $s$; furthermore, we define $N_s$ as the density of fluid species $s$ measured in the pulsar frame. The equations governing this system are \citep{levinson2005large}
\begin{equation}
\label{eq:lorentzf}
m \partial_t \left(N_s \gamma_s v_s\right) = N_s q_s E,
\end{equation}
\begin{equation}
\label{eq:sourcef}
\partial_t N_s = S/2,
\end{equation}
and
\begin{equation}
\label{eq:amperef}
\partial_t E + 4 \pi \sum_s q_s N_s v_s = 0,
\end{equation}
with $m$ again representing the pair mass, $q_s$ the species charge, $E$ the electric field, and $S/2$ the source rate. Under equivalent normalizations to those used in Section~\ref{sec:setup}, these become
\begin{equation}
    \partial_{\hat{t}} \left(\hat{N}_s \gamma_s \beta_s\right) = \hat{N}_s \frac{q_s}{e} \hat{E} ,
\end{equation}
\begin{equation}
    \partial_{\hat{t}} \hat{N}_s = 1,
\end{equation}
and
\begin{equation}
\label{eq:amperefnorm}
    \partial_{\hat{t}} \hat{E} + \frac{1}{\xi} \sum_s \frac{q_s}{e} \hat{N}_s \beta_s = 0.
\end{equation}

Except over short timescales when the pair fluid velocities reverse direction, the electron and positron fluids are highly relativistic, such that~\eqref{eq:amperefnorm} approaches
\begin{equation}
\label{eq:saw}
     \partial_{\hat{t}} \hat{E} = - \frac{1}{\xi} \sum_s \frac{q_s}{e} \hat{N}_s ;
\end{equation}
with constant or slowly changing density, this represents linear or nearly-linear change in the magnetic field, giving the sawtooth form seen in the first panel of Figure 1 of~\cite{levinson2005large}.

A kinetic description of the polar cap oscillations, presented in Section~\ref{sec:setup}, differs fundamentally from a fluid one in that it takes into account that a plasma is composed of particles of many different velocities, unlike a fluid, which has a single velocity.  Thus, the replacement presented in~\eqref{eq:saw} is not possible, and the electric field evolution is sinusoidal, with a frequency given by~\eqref{eq:freq}.  We note that to obtain a fluid-like result, i.e. one with linear oscillations of $\partial_{\hat{t}}^2 \hat{E} = 0$, we could take the limit $\left<1 /\gamma^3\right>_{+,-} \rightarrow 0$ in the frequency~\eqref{eq:freq}, which corresponds to the artificial situation of a plasma composed just of particles moving with $c$, which bears some resemblance to a fluid moving at at $c$. 

\begin{figure}
\centering
  \includegraphics[width=0.6\columnwidth]{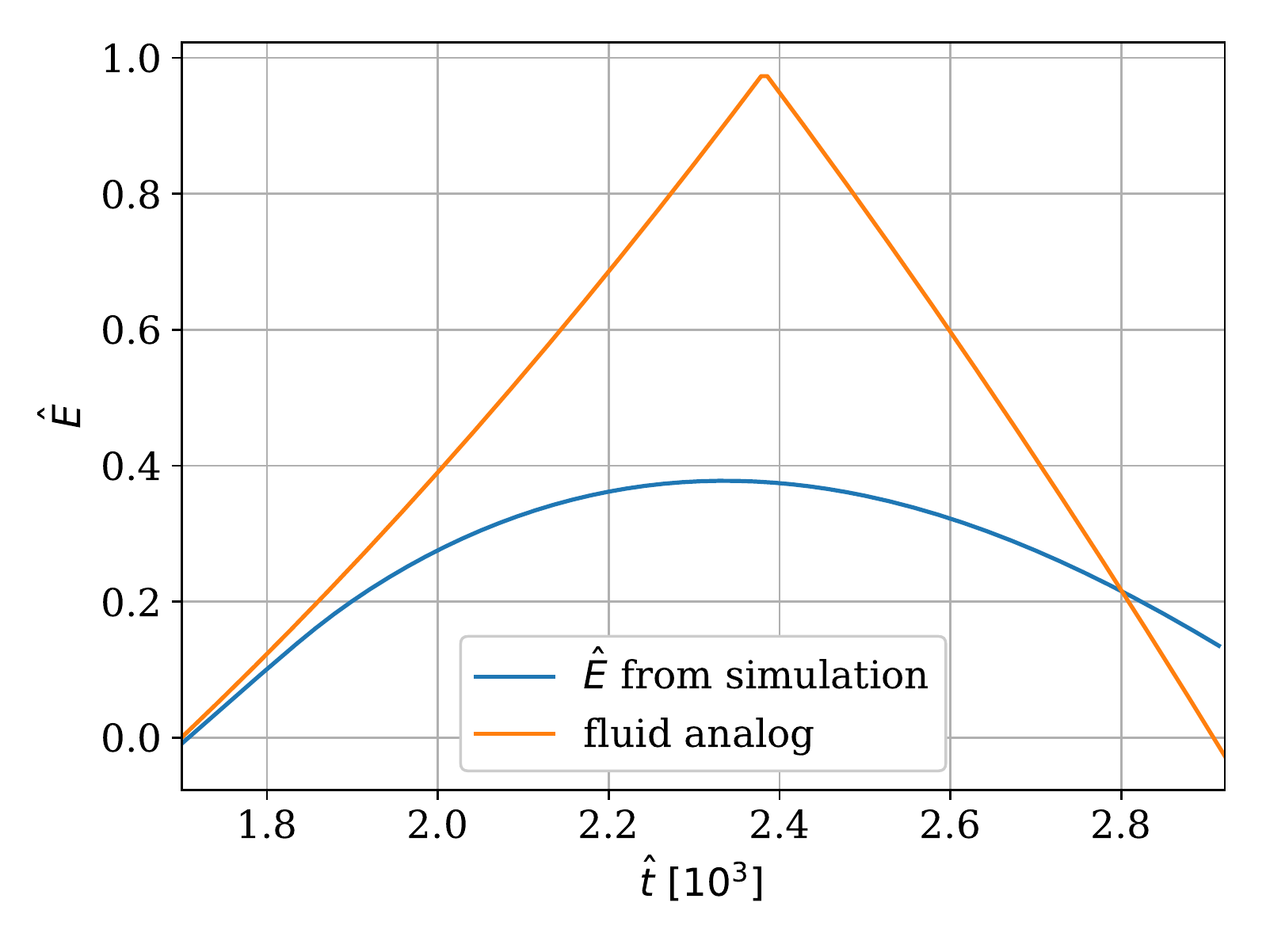}
  \caption{Comparison of electric field evolution in the first kinetic bounce shown in Figure~\ref{fig:nonlinfig} with its fluid analog.}
\label{fig:fluid}
\end{figure}

As further illustration of the difference between kinetic and fluid oscillations, we present in Figure~\ref{fig:fluid} a comparison between the first kinetic bounce shown in Figure~\ref{fig:nonlinfig} and a calculation of the fluid analog evolution of the electric field which would occur if~\eqref{eq:saw} governed the system. The difference in the two lines is significant.

\section{Estimate of scaling of $A$ and $\hat{t}_b$}
\label{sec:scaling}
In this appendix, we develop a rough model for $A$ and $\hat{t}_b$. To do this, we consider the behavior of~\eqref{eq:distnew} near the beginning of a period in which electrons are decelerated, $\hat{t}= \frac{\pi }{2\hat{\omega}}$. Specifically, consider that the contribution of new particles to the frequency is
\begin{equation}
    \hat{\omega}^2_{new} \left(\hat{t}\right) = \left(\frac{ \hat{n}_+}{\xi} \left< \frac{1}{\gamma^3}\right>_+ \right)_{new}  + \left( \frac{ \hat{n}_-}{\xi} \left< \frac{1}{\gamma^3}\right>_- \right)_{new} 
    \approx \frac{1}{\xi} \int_{u_{max} \left(\hat{t} \right) }^{ u_l }du \hat{f}_{-,new} \left(u,\hat{t} \right)\left(1+u^2\right)^{-3/2},
\end{equation}
with $\hat{f}_{-,new}$ defined in~\eqref{eq:distnew} and $u_{max} \left(\hat{t}\right)$ defined in~\eqref{eq:umax}. The frequency is dominated by contributions from the electrons because positrons are accelerated to even higher $u$ than their injection momentum $u_l$, increasing their Lorentz factor and decreasing their contribution to the frequency. 
The value of $A$ represents the height of the spike  in $\xi \hat{\omega}^2 $ that occurs just past $\hat{t}= \frac{\pi }{2\hat{\omega}}$, at the time where particles added near the zero of the electric field are pulled through $u=0$. The value of $A \hat{t}_b$ represents the area under this spike.  Specifically, we can write 
\begin{equation}
    A\sim \xi \max_{\frac{\pi}{2\hat{\omega}} < \hat{t} < \frac{3\pi}{2\hat{\omega}} } \left[ \hat{\omega}_{new}^2 \left(\hat{t} \right) \right] .
\end{equation}
and
\begin{equation}
    A\hat{t}_b \sim \xi  \int_{\pi/2\hat{\omega}}^{3\pi/2\hat{\omega}} \hat{\omega}_{new}^2 \left(\hat{t}\right)   d \hat{t}.
\end{equation}
We numerically evaluate these quantities for $\hat{E}_n\sim 1$ at several values of $\hat{\omega}$ and fit the resulting data, giving
\begin{equation}
\label{eq:A1}
    A\sim \hat{\omega}^{-0.5}
\end{equation}
and
\begin{equation}
    \hat{t}_b \sim \hat{\omega}^{-0.7}.
\end{equation}
We note from~\eqref{eq:oscfreq} that at times of zero-crossing $\hat{\omega}^2 \sim A /\xi $, and use~\eqref{eq:A1} to write  
\begin{equation}
    \hat{\omega}^2 \sim \frac{1}{\xi \hat{\omega}^{0.5}},
\end{equation}
whence,
\begin{equation}
\label{eq:A}
   A\sim \xi^{0.2}
\end{equation}
and
\begin{equation}
\label{eq:tb}
   \hat{t}_b\sim \xi^{0.3}.
\end{equation}
Simulation confirms that the height and width of the spikes increase slowly with $\xi$.

\section{Effect of finite temperature}
\label{sec:temp}
\begin{figure*}
\begin{subfigure}
\centering
\includegraphics[width=0.5\columnwidth]{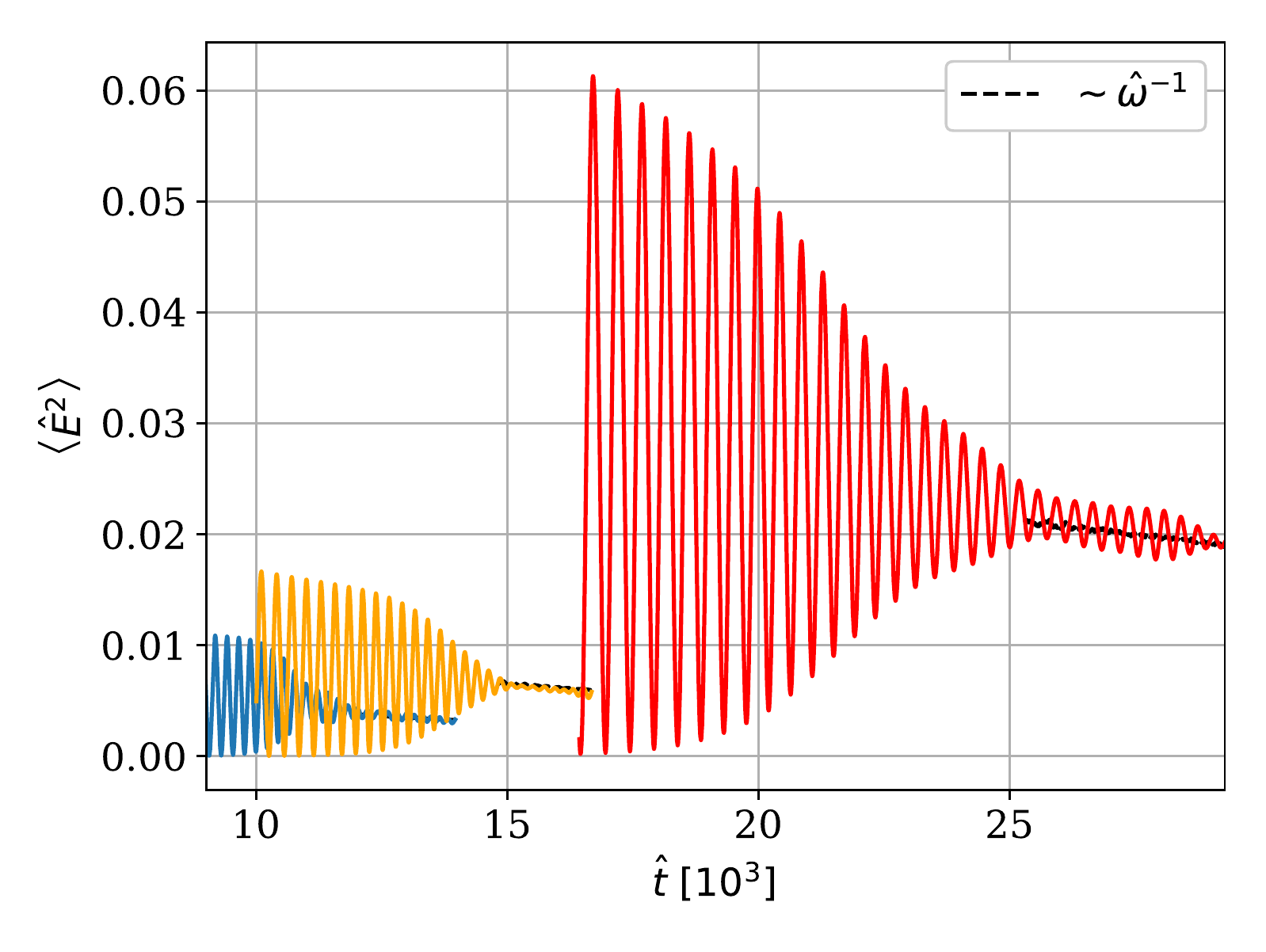}
\end{subfigure}%
\begin{subfigure}
\centering
\includegraphics[width=0.5\columnwidth]{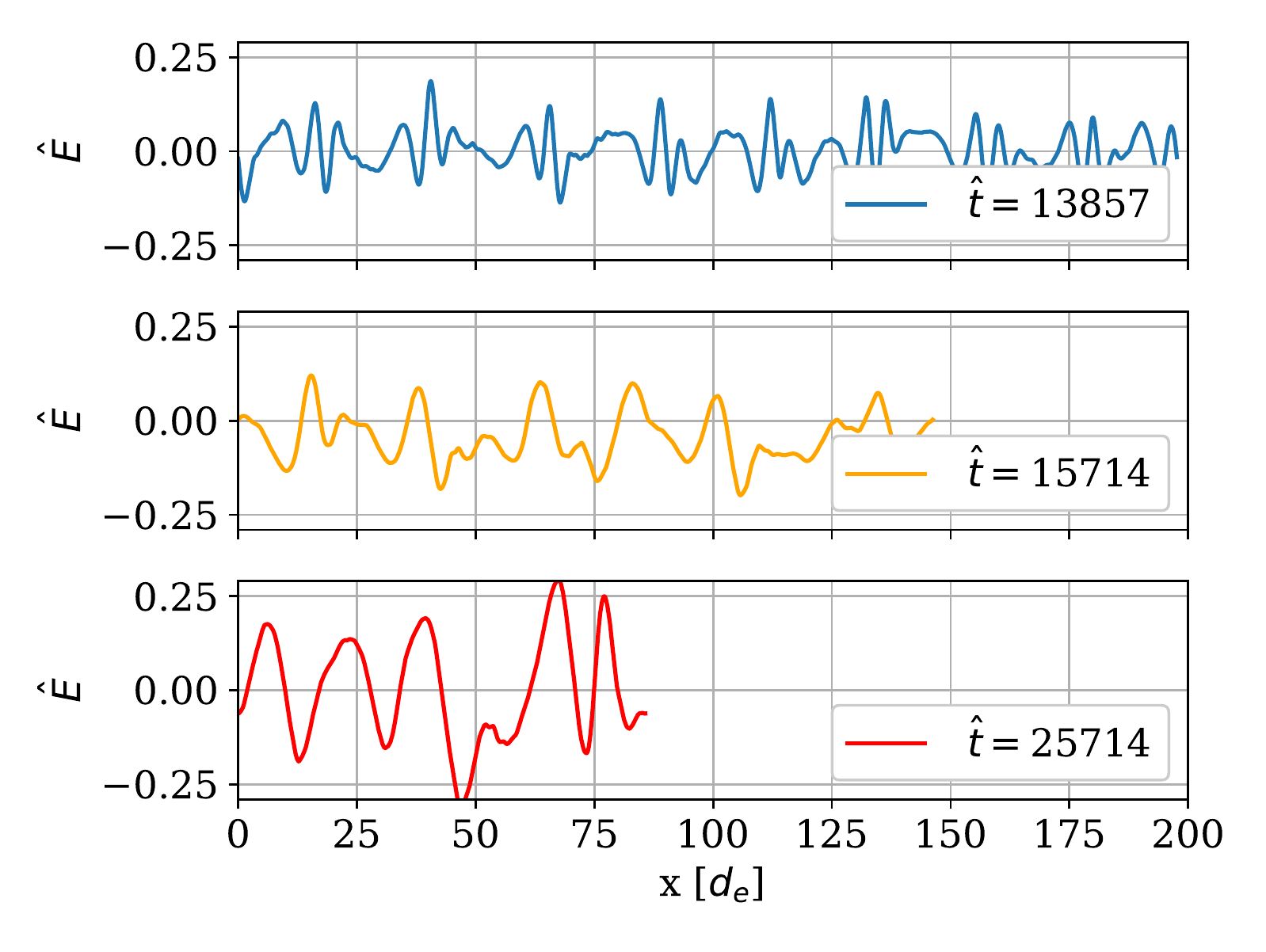}
\end{subfigure}
\caption{Comparisons of important quantities for different temperatures: $0.1$ mc$^2$ (blue), $1$ mc$^2$ (orange), and $10$ mc$^2$ (red). Note that, in the right plot, the x-axis extends to different values of $x$ in units of skin depth, even though the simulation box size is the same, because higher temperature simulations have a larger skin depth.}
\label{fig:tempeff}
\end{figure*}

All analysis after~\eqref{eq:vlasov} assumes negligible injection temperature; our simulations use a small injection temperature of $T = 0.1 \, mc^2$, as described in Section~\ref{sec:sim}.  However, as noted in Section~\ref{sec:setup}, newly injected pairs in fact have a thermal spread of roughly $T \gtrsim m c^2$.  To examine the effect of this thermal spread, we run two more cases of our $\xi = 2.8 \times 10^6$ simulation with injection temperatures of $T= 1 \, m c^2$ and $T= 10 \, m c^2$. The effect of finite injection temperature on these simulations can be seen in Figure~\ref{fig:tempeff}. The left figure, which depicts the quantity also shown in Figure~\ref{fig:linstage}, shows that higher temperature simulations have a lower frequency and a higher saturation amplitude. In addition, the fragmentation into finite $k$ modes occurs later in the higher temperature simulation. The right plot in Figure~\ref{fig:tempeff} displays the spatial profile of $\hat{E}$ at a time late in the discharge, after the simulation fragments into finite-$k$ modes, and should be compared to the quantities in Figure~\ref{fig:frag}, showing that higher temperature simulations have slightly larger wavelength modes.


\bibliography{sample631}{}
\bibliographystyle{aasjournal}



\end{document}